%% file: main.tex
\documentclass[sigconf]{acmart}

\usepackage{xspace}
\usepackage{bm}
\usepackage{subfigure}
\usepackage{graphicx}

\newtheorem{remark}{Remark}

\newcommand{\ie}{\emph{i.e.,}\xspace}

\newcommand{\eg}{\emph{e.g.,}\xspace}

\newcommand{\wrt}{\emph{w.r.t.}\xspace}

\newcommand{\model}{COUPA}

\AtBeginDocument{%
  \providecommand\BibTeX{{%
    \normalfont B\kern-0.5em{\scshape i\kern-0.25em b}\kern-0.8em\TeX}}}

\setcopyright{acmlicensed}
\copyrightyear{2023}
\acmYear{2023}
\acmISBN{978-1-4503-9408-6/23/07}
\acmDOI{10.1145/3539618.3591828}

\acmConference[SIGIR '23]{Proceedings of the 46th International ACM SIGIR Conference on Research and Development in Information Retrieval}{July 23--27, 2023}{Taipei, Taiwan}
\acmBooktitle{Proceedings of the 46th International ACM SIGIR Conference on Research and Development in Information Retrieval (SIGIR '23), July 23--27, 2023, Taipei, Taiwan}
\acmPrice{15.00}

\begin{document}

\title{{\model}: An Industrial Recommender System for Online to Offline Service Platforms }

\author{Sicong Xie$^*$, Binbin Hu$^*$, Fengze Li, Ziqi Liu, Zhiqiang Zhang, Wenliang Zhong,  Jun Zhou$^\#$}
\thanks{$^{*}$ Equal contributions.   }
\thanks{$^\#$ Corresponding author.}
\affiliation{%
\institution{Ant Group, Hangzhou, China}
}
\email{
{sicong.xsc, bin.hbb, fengze.lfz, ziqiliu, lingyao.zzq, yice.zwl, jun.zhoujun}@antfin.com
}

\begin{abstract}


Aiming at helping users locally discovery retail services (\eg entertainment and dinning), Online to Offline (O2O) service platforms  have become popular in recent years, which greatly challenge current recommender systems. With the real data in Alipay, a feeds-like scenario for O2O services, 
we find that recurrence based temporal patterns and position biases commonly exist in our scenarios, which seriously threaten the recommendation effectiveness.
To this end, we propose \textbf{\model}, an industrial system targeting for characterizing user preference with following two considerations:
(1) Time aware preference: we employ the continuous time aware point process equipped with an attention mechanism to fully capture temporal patterns for recommendation.
(2) Position aware preference: a position selector component equipped with a position personalization module is elaborately designed to mitigate position bias in a personalized manner.
Finally,  we carefully implement and deploy {\model} on Alipay with a cooperation of edge, streaming and batch computing, as well as a two-stage online serving mode, to support several popular recommendation scenarios.
We conduct extensive experiments to demonstrate that {\model} consistently achieves superior performance and has potential to provide intuitive evidences for recommendation.

\end{abstract}







\maketitle

\input{1-sec-intro}
\input{2-sec-data}
\input{8-sec-pre}
\input{3-sec-model}
\input{4-sec-sys}
\input{5-sec-exp}
\input{6-sec-rel}
\input{7-sec-con}

\bibliographystyle{ACM-Reference-Format}
\bibliography{references}



\end{document}

%% file: 1-sec-intro.tex
\section{Introduction}

Recommender systems, which aim at matching user interests, are playing vital role in various online websites and mobile applications, including Taobao~\cite{gong2020edgerec}, Youtube~\cite{covington2016deep} and Google Play~\cite{cheng2016wide}. To fully mine user preference in online services, numerous methods are proposed to learn the effective matching function between target users and items with subtle feature engineer~\cite{xiao2017attentional,lian2018xdeepfm,guo2017deepfm} or abundant historical records~\cite{zhou2018deep,zhou2019deep,hidasi2015session,kang2018self,hou2022core}. Although these methods achieve superior performance to some extent, we argue that they are unsuitable to recommendation scenarios for Online to Offline (O2O) service platforms (\eg Meituan~\footnote{https://about.meituan.com/en}, Grubhub~\footnote{https://www.grubhub.com/} and Uber Eats~\footnote{https://www.ubereats.com/}), where recurrence based temporal patterns are ubiquitous.

\begin{figure}
    \centering
    \includegraphics[width=8.0cm]{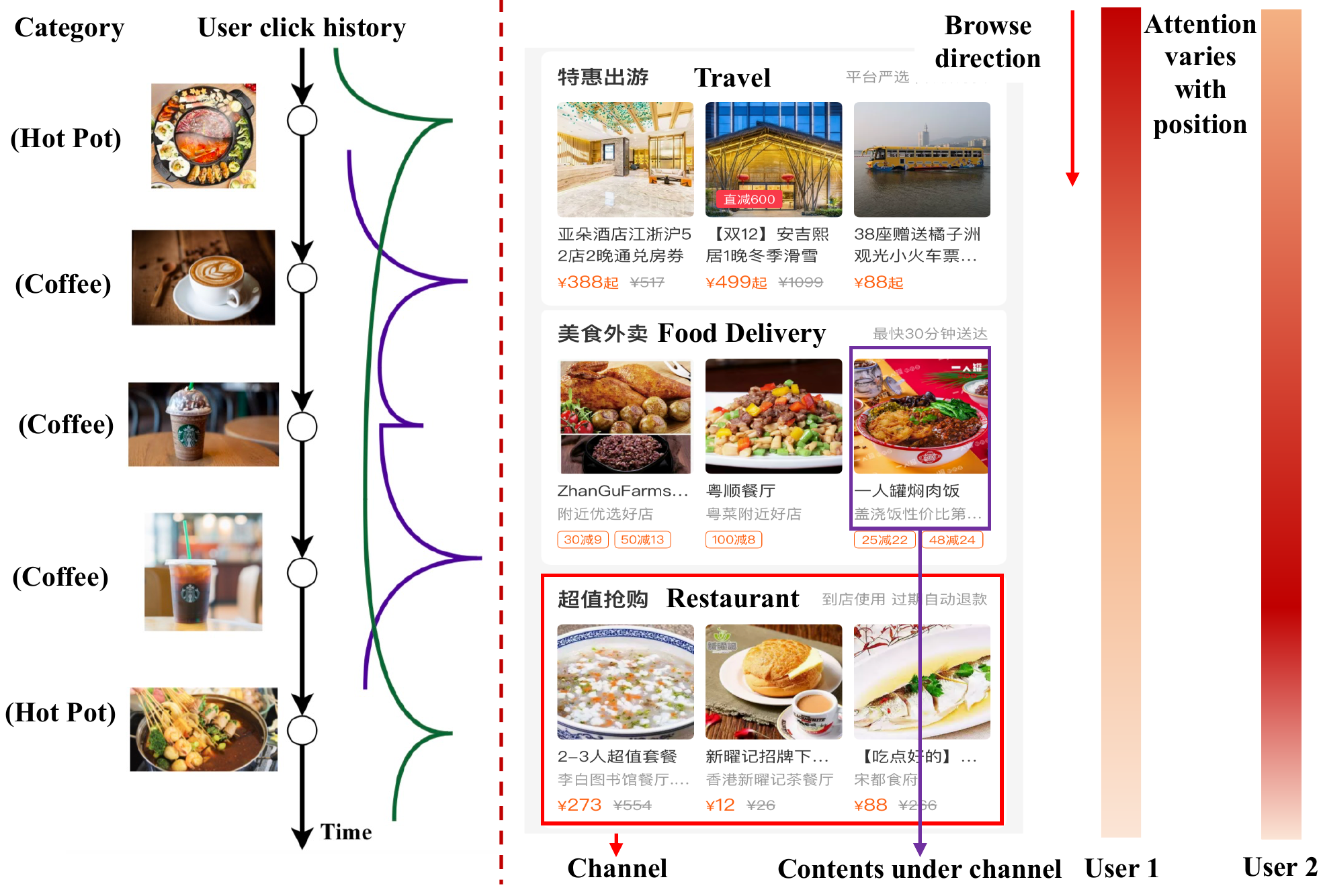}
    \caption{The illustrated example of recommendation scenarios in Alipay. The left part shows the user's evolving demands over time, while the right part implies varied attention distributions with browsing for different users. }
    \label{fig-alipay}
\end{figure}

In the case of Alipay~\footnote{https://www.alipay.com/}, whose goal is to guide user for locally discovering retail services including entertainment, travelling, delivery and other services, there is a clear and strong need to capture dynamic preference over time for recommendation. Concretely, we illustrate an example in Figure~\ref{fig-alipay}. There are numbers of channels (\eg ``Food Delivery'' and ``Travel'' channel) in our scenarios, each of which represents a certain business field. Besides, several items are displayed under each channel, which are called contents of channel. Taking the contents of ``Food Delivery'' channel~\footnote{Obviously, similar findings can also be revealed for other channels (\eg ``Travel'' channel).} as an example (as shown in the left part of Figure~\ref{fig-alipay}), users may be interested in daily necessities (\eg ``Coffee'') or weekly intents (\eg 
``HotPot''), leading to different click demands at different times.
On the other hand, as shown in the right part of Figure~\ref{fig-alipay}, since the feeds-like styles are widely employed in the recommendation scenarios of O2O service platforms, users typically scan the screen of mobile phone from top to bottom, while their attention distributions are indeed distinct. That is, users may prefer clicking items with certain positions regardless of the relevance,
which brings so-called personalized position biases to  hurt the recommendation effectiveness.



In light of the above observations, we believe it is of crucial importance to design an industrial recommender system for O2O service platforms, centering on characterizing user preference from time and position bias aware perspectives simultaneously. The idea is appealing, while the solution is challenging in real recommendation scenarios, which is summarized as follows:   
(\textbf{C1}) \emph{Time aware preference modelling: how to effectively capture recurrence based preferences for users over continuous time?} 
Recurrence based temporal patterns are ubiquitous in recommendation scenarios for local services. For example, a certain user may periodically visits specific malls and enjoy specific delicacies (\eg Hot Pot) in a reasonable period (\eg weekend). Therefore, explicitly capturing such temporal patterns over continuous time is potential to benefit for the preference characterization in real recommendation scenarios.
(\textbf{C2}) \emph{Position aware preference modelling: how to deal with user's preferences for positions and build a debiasing model in a personalized way?} 
On account of the feeds-like exhibition in our scenarios and users' browsing habits, position biases widely exist in the behavior data. Moreover, users have different preferences towards each position.
For example, most users prefer clicking items ranked first whereas some users have preferences for other certain positions (\eg the second and even last position). Blindly fitting the observational behavior data without considering the inherent position biases may seriously deteriorate the recommendation effectiveness.
(\textbf{C3}) \emph{System design: how to devise the base system to support the complicated temporal interaction scenarios?}
Our temporal interaction scenario put forward the urgent request of low delay for system design since our approach is expected to have awareness of user real-time behaviors for online serving. Moreover, building the complete chain to balance the trade off between data storage and timely feedback
as well as guaranteeing low delays for online serving also remain challenging.

To this end, we propose \textbf{\model}, a novel \underline{CO}ntinuo\underline{U}s time and \underline{P}osition bias \underline{A}ware recommender system targeted for O2O service platforms (\eg Alipay). 
Specifically, we borrow the idea of functional time encoding, a promising way to embed timestamp into vector space, and propose a novel continuous time aware point process with an attention mechanism to explore the excites of inhabits from previous interaction  records in continuous time (\textbf{C1}).
On the other hand, inspired by the idea of knowledge transfer, a position selector component, cooperated with a position personalization module is elaborately designed to perform personalized position debiasing in our scenarios (\textbf{C2}). 
To satisfy the requirements of online serving, the implementation and deployment are carefully designed to not only collect user behaviors as well as corresponding positions in a real-time manner, but also perform efficient online inference in a two-stage mode (\textbf{C3}).

In sum, we make the following contributions:
\begin{itemize}
    \item We analyze and highlight the importance of capturing time and position aware preferences in Alipay, which is potentially generalized to the universal recommendation scenarios for O2O service platforms (\eg Meituan, Grubhub and Uber Eats).

    
    \item We propose {\model}, which is not only capable of modeling continuous time aware preference for recommendation, but also
    well deal with position bias in a personalized manner.
    
    \item System deployment and corresponding implementation details are uncovered, where edge, streaming and batch computing are jointly employed for capturing the real-time user behaviors, and a two-stage online serving mode is also carefully designed.
    
    \item We conduct extensive offline and online experiments, demonstrating that  {\model} consistently achieves superior performance and provides intuitive evidences for recommendation. 
\end{itemize}

%% file: 2-sec-data.tex
\section{Observations on Real Data}
In this section, we give an intuitive analysis for user preference \wrt temporal pattern and position existed in recommendation scenarios for O2O platform, with Alipay as an example.
All of the real data are collected within one month, consisting of user interactions with channels and contents under each channel.


\begin{figure}[t]
\centering
\subfigure[]{
\includegraphics[width=40mm,height=30mm]{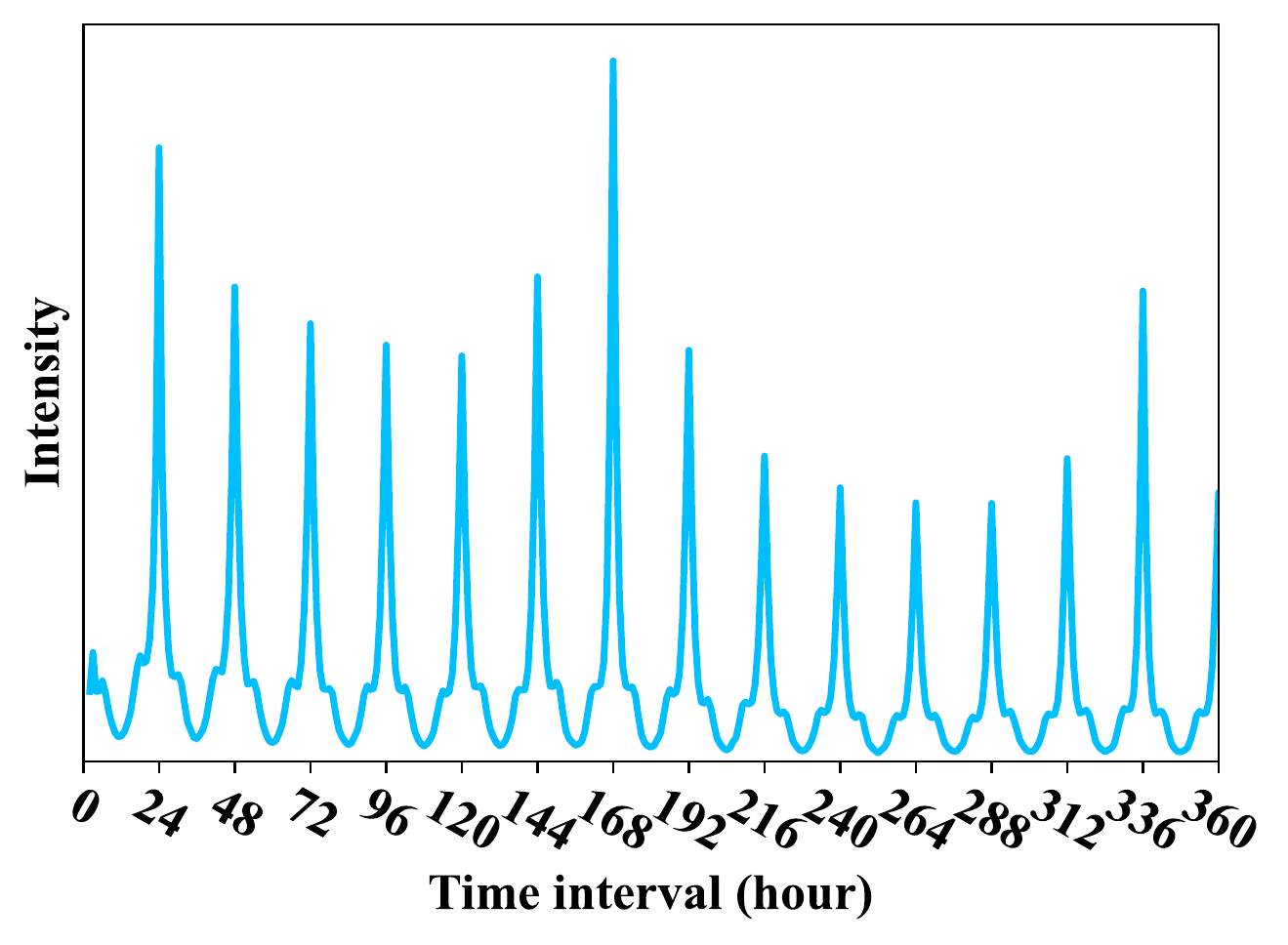}
}
\subfigure[]{

\includegraphics[width=40mm,height=30mm]{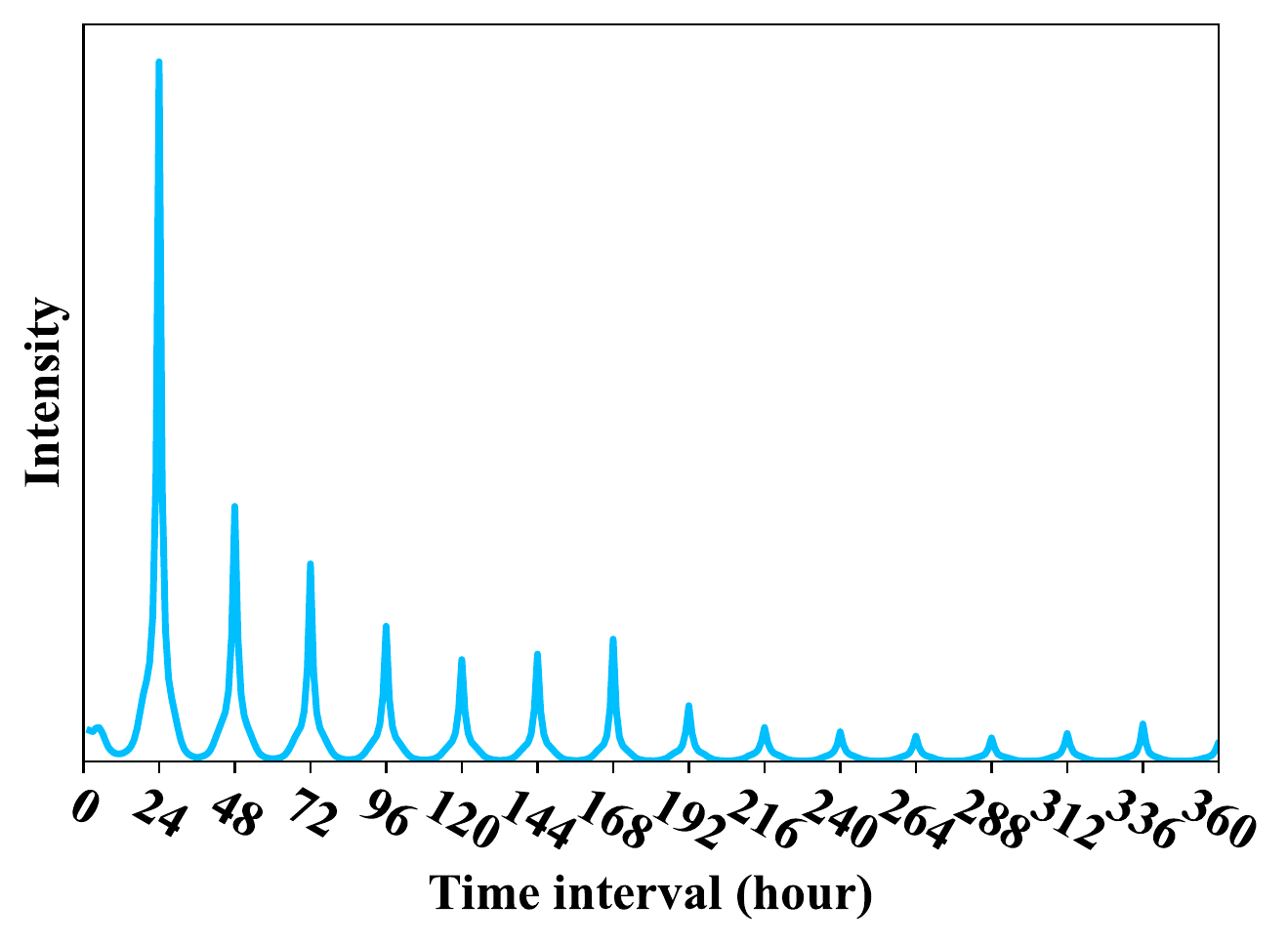}
}
\subfigure[]{

\includegraphics[width=26mm,height=30mm]{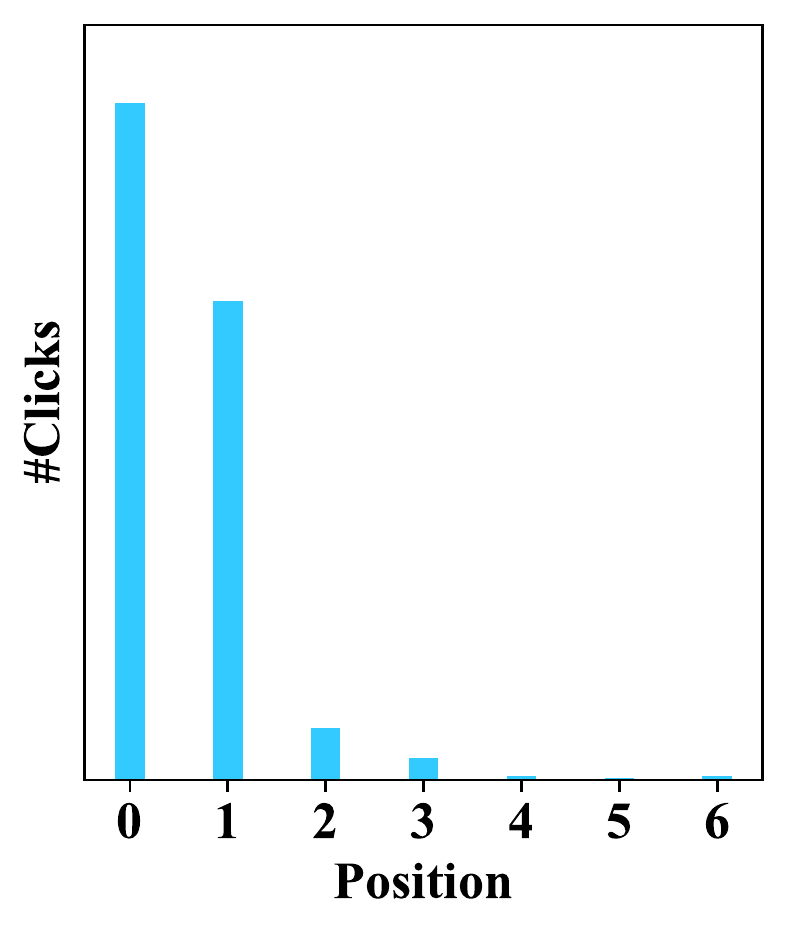}
}
\subfigure[]{

\includegraphics[width=26mm,height=30mm]{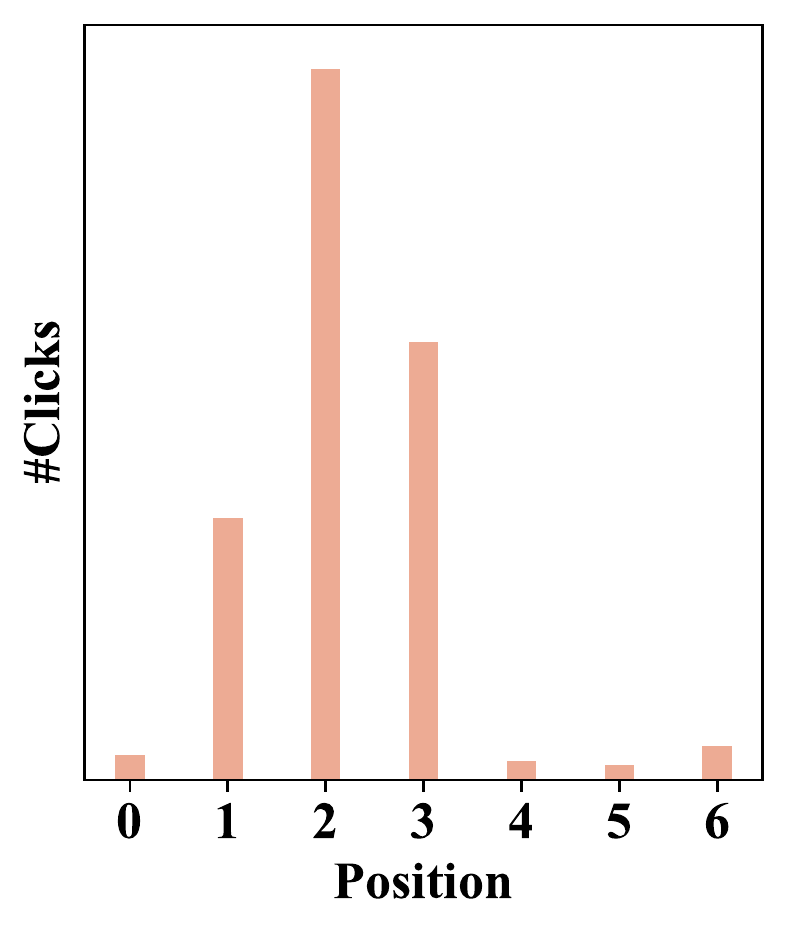}
}
\subfigure[]{

\includegraphics[width=26mm,height=30mm]{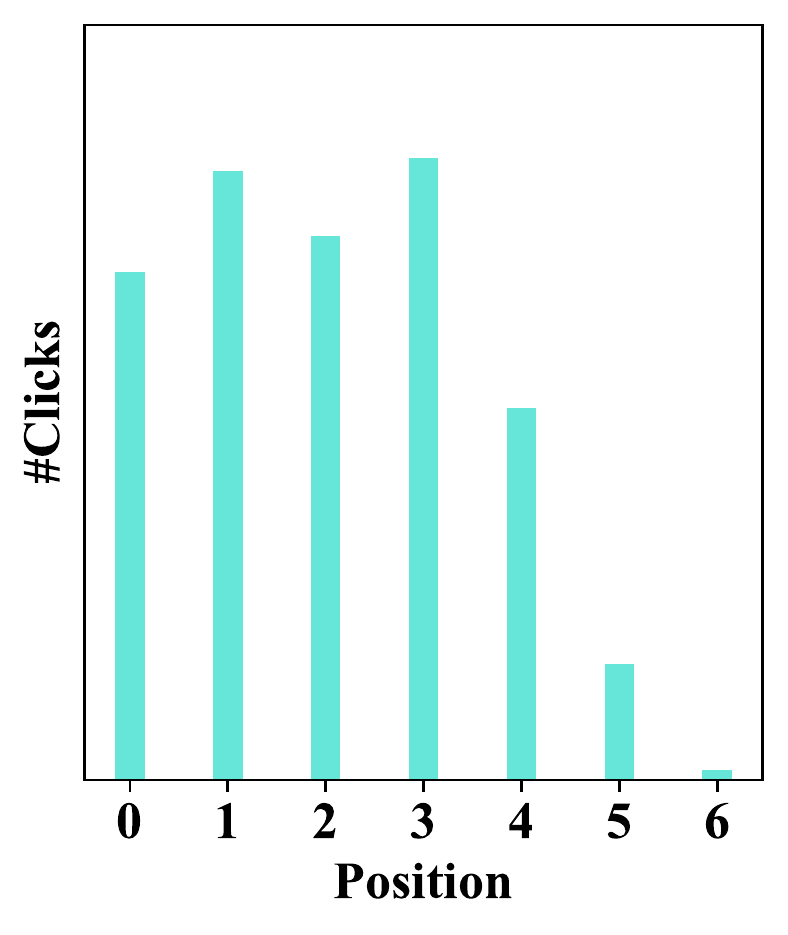}
}
\caption{(a), (b): The temporal pattern analysis on ``Hot Pot'' and ``Coffee'', respectively. (c), (d), (e): The  analysis for position bias on channels for ``User type 1'', ``User type 2'' and ``User type 3''. The smaller the number of $x$-axis, the higher the position. \label{fig-data}}
\end{figure}

\textbf{Temporal Pattern.} As mentioned above, recurrence based temporal patterns are widely existed in recommended scenarios for O2O services. Since users may be interested in  daily necessities or weekly intents, which implies that the reasons of users accessing O2O service platform (\eg Alipay) are periodical. With the real data, we aim at digging out the willingness of users to purchase items under the same categories \wrt time interval.

We respectively select the categories  ``Hot Pot" and ``Coffee'' and plot the corresponding intensities (\ie the number of purchase) against time intervals in Figure~\ref{fig-data} (a) and (b). Here, $t = 0$ refers to the timestamp of first purchase. From results in Figure~\ref{fig-data} (a), we observe that the intensity reaches local peak when $t = 24 \cdot n (n = 1, 2, ...)$. Moreover, among these peaks, the intensity is more large when  $t = 7 \cdot 24 \cdot n (n = 1, 2)$. These observations show that the intention of purchasing ``Hot Pot'' has  strong daily and weekly periodicity. On the contrary, the intensity under the ``Coffee'' achieve a peak with 24 hours after the first purchase, whereas it  sharply declines as time passes. Both cases reveal that user behaviors have strong temporal patterns \wrt different categories of items in our recommendation scenarios, which is vital information that should be fully considered in our approach.



\textbf{Position bias.}
Click distributions are always distinct over positions for different users in feeds-like scenarios of O2O platform, which brings in personalized bias in modelling caused by that user feedback data is observational rather than experimental.

We conduct a micro-view analysis based on 3 types of users with different click distributions over positions in Figure~\ref{fig-data} (c), (d), (e). Specifically, ``User type 1'', whose behavior is the most common case, prefers clicking items with higher positions (\ie Position ``0'' and ``1''), whereas ``User type 2'' is more likely to click position ``2'' and ``3''. As for ``User type 3'', who treat each position almost equally, their preferences are more stable. 
In sum, we conclude that position bias is not only dependent, but also personalized, which has not been sufficiently explored in previous studies.

%% file: 8-sec-pre.tex
\section{Preliminary}

\textbf{Functional time encoding} aims to find a mapping  $\Phi: \mathcal{T} \rightarrow \mathbb{R}^d$  from time domain $\mathcal{T}$ to $d$-dimensional vector space, targeting for preserving evolving nature of user interest/intent. Intuitively, the temporal pattern related to the timespan between any  two timestamps $t_1, t_2 \in \mathcal{T}$ can be denoted as inner product between their functional encodings, \ie $\langle \Phi(t_1), \Phi(t_2) \rangle$. Therefore,  we formulate above temporal patterns with a translation-invariant kernel $\mathcal{K}$ with $\Phi$  as the mapping function associated with $\mathcal{K}$.

Suggested by the Mercer's Theorem~\cite{minh2006mercer,xu2019self,humerit}, we formulate the mapping function $\Phi$ with frequency parameter $\omega$ as follows:
\begin{equation}
\label{eq_k_w}
    t \mapsto \Phi_{\omega}^{\mathcal{M}}(t):=[\sqrt{c_1},..., \sqrt{c_{2j}}cos(\frac{j\pi t}{\omega}),  \sqrt{c_{2j+1}}sin(\frac{j\pi t}{\omega}), ...]^T.
\end{equation}
With the help of the nice truncation properties provided by such a  Fourier series-like form, we truncate above mapping function $\Phi_{\omega}^{\mathcal{M}}(t)$ as $\Phi_{\omega, d}^{\mathcal{M}}(t)$. Subsequently, by concatenate multiple truncated periodic mapping function with the frequency set $\{\omega_1,...,\omega_k\}$, we represent time with functional encoding as:
\begin{equation}
\label{eq_k_w}
    t \mapsto \Phi_d^{\mathcal{M}}(t) :=  [\Phi_{\omega_1, d}^{\mathcal{M}}(t)||...||\Phi_{\omega_k, d}^{\mathcal{M}}(t)]^T.
\end{equation}

\textbf{Temporal point process} has been commonly adopted to model dynamic in sequences, which is a kind of stochastic process that generates a list of discrete events at different times, denoted as  $\mathcal{S} = \{v_i, t_i\}_{i = 1}^{n}$ where $v_i \in \mathcal{V}$ is the type (\ie item in our study) of $i$-the event and $t_i \in \mathcal{T}$ is the timestamp of the $i$-th event. The process is well characterized via the conditional intensity function $\lambda^*(t) := \lambda(v, t|\mathcal{H}_t)$, which makes prediction for next arrival time $t$ based on the history $\mathcal{H}_t = \{(v_i, t_i) \in \mathcal{S} | t_i < t\}$. Here, the $*$ symbol reminds us of dependence on $\mathcal{H}_t$. 
The core of the temporal point process is to design the conditional intensity function $\lambda^*(t)$ for capturing various interests. As consequence, a series of typical point processes equipped with well-designed intensity functions are proposed, including Possion process, Hawkes process and recently proposed  fully neural network based proint process (FullyNN).

Maximum likelihood estimation (MLE) is commonly adopted to learn the parameters of temporal point process, and the corresponding  likelihood over a time interval $[0, T]$ is given by :
\begin{equation}
    \mathcal{L}_{\mathcal{T}} = \sum_{i = 1}^{n}\log{\lambda_{v_i}(t_i)} - \int_{0}^{T}{\lambda(\tau)d\tau},
\end{equation}
where the first term models the sum of log-intensity functions of past events and the second term is the log-likelihood of infinitely many non-events, where the negative sampling strategy is always adopted.

%% file: 3-sec-model.tex
\section{The Proposed {\model} \label{sec-model}}
\begin{figure*}
    \centering
    \includegraphics[width=18cm]{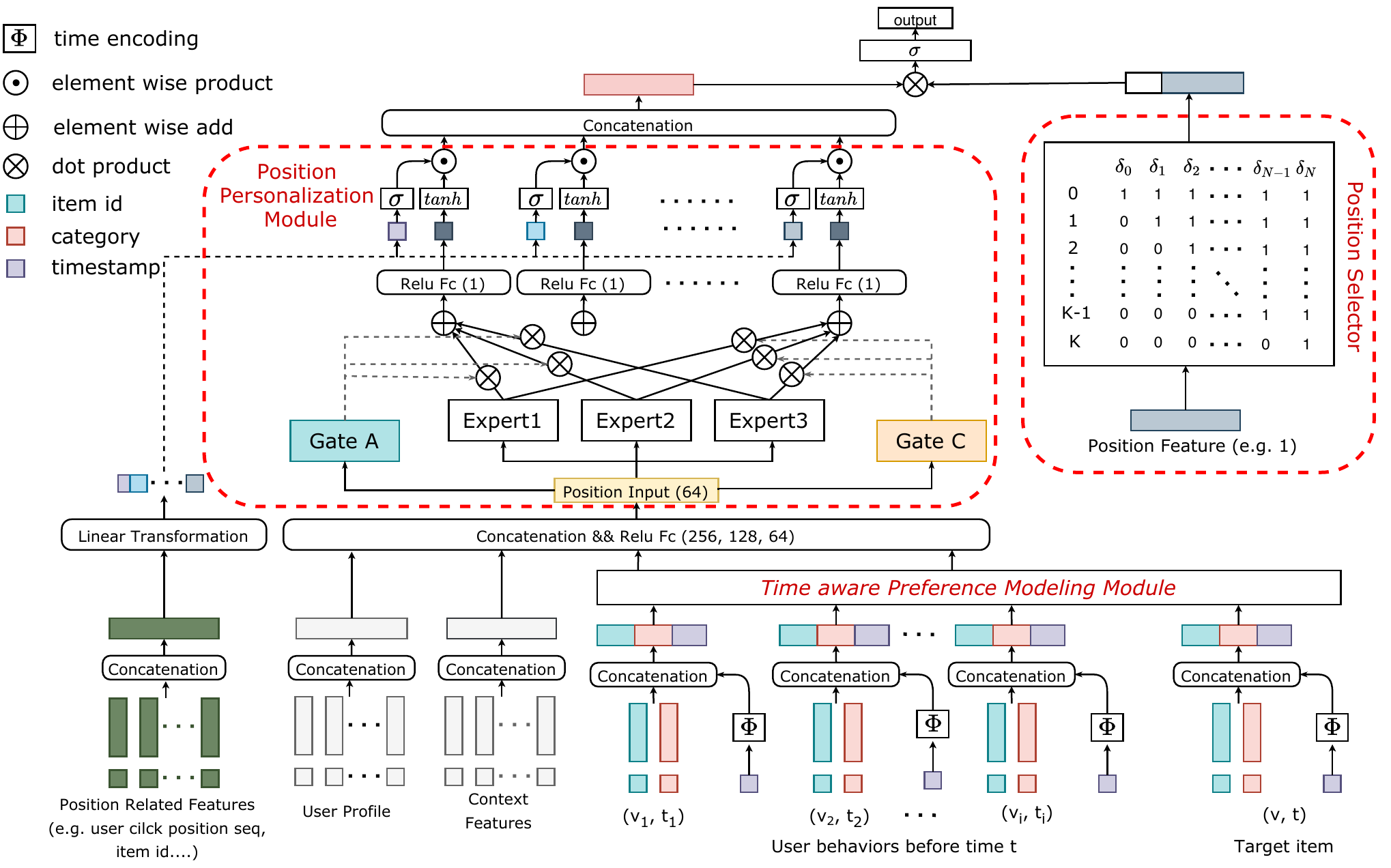}
    \caption{The overall architecture of the proposed {\model}.}
    \label{fig-model}
\end{figure*}

In this section, we present {\model}, a novel \underline{CO}ntin\underline{U}ous time and \underline{P}osition \underline{A}ware recommender system
for O2O platforms,
whose model part mainly focuses on  user preferences towards temporal patterns (\ie time aware preference modelling) and positions (\ie position bias aware modelling).
The overall architecture of {\model} is illustrated in Figure~\ref{fig-model}. 

Before the elaboration of the model design for {\model}, we give a brief introduction for the inputs, which clearly consist of position related features (\eg the sequence of user click positions and the corresponding item ids), user profile, context feature and user behaviors (including item ids and category ids) from left to right in Figure~\ref{fig-model}. Following common strategies adopted in \cite{guo2017deepfm,lian2018xdeepfm}, we transform the involved features into low-dimensional representations (called embedding) by look-up operation, concatenation and multi-layer perceptron successively. At last, given a user $u$ and an item $v$, we represent the embedding for position related features as $\mathbf{p}_{u,v}$, user profile as $\bm{e}_u$, context features as $\mathbf{c}_{u,v}$ and item profile (\ie item id and category id) as $\bm{e}_v$, respectively.

\subsection{Time aware Preference Modelling}
Generally, user's preferences are affected by the items they have already interacted (\eg click and purchase) with and the effects evolve as time passes. Due to the fact that modelling such preferences is highly time sensitive, we firstly represent the timestamps into a continuous, higher dimensional space for preserving the temporal patterns with the help of the functional time encoding. 
Subsequently, we propose continuous time aware point process module to summarize influence from historical interaction records for estimating the likelihood of the target item, where a novel continuous time aware attention mechanism is introduced for locating important/relevant items in continuous time.  

\subsubsection{Continuous Time aware Attention Mechanism.} To effectively model the target user's periodic preferences derived from his/her historical behaviors over continuous time spaces, we introduce the continuous time aware attention mechanism to chronologically weigh various underlying preferences for historical interactions \wrt the target item. 
We devise it upon the masked self-attention architecture~\cite{vaswani2017attention}
to learn the adaptive weights over continuous time conditioned on the involved user behaviors and target item. 
Formally, given a triplet $\langle u, v, t \rangle$, we denote the historical interaction records  before time $t$ as $S_{u,v} = \{(v_i, t_i) | t_i < t\}$. Following the original self-attention  mechanism~\cite{vaswani2017attention}, we obtain the temporal sequence matrix at time $t$ to take  account of the relationships of user behaviors $S_{u, v}$ and target item $v$.
\begin{equation}
\begin{split}
    \mathbf{Z}_{u,v}(t) =  & [\bm{e}_{v_0}\ ||\ \Phi_d^{\mathcal{M}}(t - t_0);\ ...; \\
                           & \bm{e}_{v_{N - 1}}\ ||\ \Phi_d^{\mathcal{M}}( t - t_{N-1}); \bm{e}_{v}\ ||\ \Phi_d^{\mathcal{M}}(0)]^T, 
\end{split}
\end{equation}
where ``||'' denotes concatenation operation, $N = |\mathcal{S}_u|$ is the length of user behaviors and $\bm{e}_*$ is the original embedding for items.  Then, we produce the final representation that summarizes the influence of user behaviors using the scaled dot-product attention, which is formulated as:
\begin{equation}
    \mathbf{h}_{u,v}(t) = \text{softmax}(\frac{\mathbf{Q}_{u,v}(t)\mathbf{K}_{u,v}^T(t)}{\sqrt{d}})\mathbf{V}_{u,v}(t),
\end{equation}
where $\mathbf{Q}_{u, v}(t), \mathbf{K}_{u, v}(t)$ and $\mathbf{V}_{u, v}(t)$ respectively denotes the ``query'', ``key'' and ``value' matrix \wrt user $u$ and item $v$ in time $t$, which are linear projections of the  temporal sequence matrix $\mathbf{Z}_{u,v}(t)$:  $\mathbf{Q}_{u,v}(t) = \mathbf{Z}_{u,v}(t)\mathbf{W}_{Q}$, $\mathbf{K}(t) = \mathbf{Z}_{u,v}(t)\mathbf{W}_{K}$ , $\mathbf{V}_{u,v}(t) = \mathbf{Z}(t)_{u,v}\mathbf{W}_{V}$, where $\mathbf{W}_{Q}, \mathbf{W}_{K}$ and $\mathbf{W}_{V}$ are weight matrices for linear projection.

\subsubsection{Continuous time aware point process.}
Here, we aim to estimate the likelihood of the target item $v$ based on the historical interaction records $\mathcal{S}_{u,v}$ through continuous time aware point process, whose major role is to construct conditional intensity function $\lambda_v(t|\mathcal{S}_{u,v})$. For convenience,  we rewrite above intensity function as $\lambda_v(t|\mathcal{S}_{u,v}) = \psi_v(t - t_i|\mathbf{h}_{u,v})$, where $\psi_v(\cdot)$ is a non-negative function, commonly implemented as the exponential function in previous works. Instead of formulating such a specific functional form, which only models the exponential effects (decrease or increase) of historical behaviors toward the target item, following the idea in ~\cite{omi2019fully}, we exploit a more complex way to enhance the model capability. That is, we directly modelling the cumulative intensity function , which can be differentiated for the final intensity function. 
\begin{equation}
    \Psi_{u, v}(\tau|\mathbf{h}_{u,v}) = \int_{0}^{\tau}\psi(s|\mathbf{h}_{u, v})ds,
\end{equation}
where $\tau = t - t_i$ denotes the interval since the last interaction. Obviously, we adopt an intensity-free formulation to model the user's time aware preference towards target items, which is more suitable to complex scenarios in real-world applications. Due to the ability of modelling non-linear functions, we implement the cumulative intensity function $\Psi_{u, v}(\cdot)$ with a feed-forward neural network as follows:
\begin{equation}
\begin{split}
    &\Psi_{u, v}(\tau|\mathbf{h}_{u,v})  = g(\mathbf{W}_L...g(\mathbf{W}[\bm{e}_u\ ||\ \mathbf{h}_{u,v}\ ||\  \tau] + \mathbf{b}_1) + b_L), \\
    &s.t. \ \ \ \ \mathbf{W}_1, ..., \mathbf{W}_L \succeq 0, b_1, ..., b_L \geq 0.
\end{split}
\end{equation}
Here, $\bm{e}_u$ is the user embedding derived from his/her original profile features and $g(\cdot)$ is the ReLU activation function. At last, we obtain our final intensity function as follows, 
\begin{equation}
    \lambda^*_v(t) = \psi_v(\tau|\mathbf{h}_{u,v}) = \frac{\partial \Psi_{u, v}(\tau|\mathbf{h}_{u,v})}{\partial \tau}.
\end{equation}

\subsection{Position Bias aware Modelling}
Through the above intuitive analysis, user's preference of clicking at certain positions may bring in personalized position bias, which may deteriorate the recommendation effectiveness without consideration. Modelling user's preference for positions is important for debiasing, whereas previous works always take the simple assumption that each position is dependent, which ignores the heavy affects of personalized features related to users. To fill this gap, we aims at performing position debiasing in a personalized manner, where a position selector component equipped with a position personalization module is elaborately designed.



\subsubsection{Position Personalization Module.}
Since positions differ from each other based on the features of specific users and items, the position uplifts in different positions can be regarded as multiple tasks, whose differences and relations can be naturally captured by the Multi-gate Mixture-of-Experts (MMoE)~\cite{ma2018modeling}. Specifically, following the idea of \cite{ma2018modeling}, we formulate so-called $k$-th position uplift for user $u$ and item $v$ as follows:

\begin{equation}
    \hat{\delta}_{u,v,k}=ReLU(\bm{v}\sum_{i=1}^{n}\text{softmax}(\mathbf{W}_{k}\mathbf{x}_{u,v})_{i}\cdot{f_{i}(\mathbf{x}_{u,v})}+b),
\end{equation}
where $\hat{\delta}_{u,v,k}$ refers to the position uplift in position $k$, $f_{i}(\cdot)$ refers to the output of $i$-th experts, $n$ is the number of experts,  $\mathbf{W}_k \in {\mathbb{R}^{d\times{d}}}$ is a trainable matrix for the $k$-th position, $\bm{v}$ and $b$ are also trainable parameters and $\mathbf{x}_{u,v}$ is generated by a 3-layer fully connected neural network with the input of profile embedding for target user (\ie $\mathbf{e}_u$), context embedding for target user and items (\ie $\mathbf{c}_{u, v}$) and continuous time aware embedding for target user and item at current time (\ie $\mathbf{h}_{u,v}(t)$).

To further enhance personalization, we incorporate a few position related features (\eg{user click position sequence, item id}) and design a Gated Linear Units (GLU)~\cite{dauphin2017language} block to control the information passed from features and the position uplift from MMoE block. Specifically, a linear transformation is applied for the position related features to obtain the gate units $\epsilon_{k}$ for $k$-th position. Then, the final position uplift can be calculated as:
\begin{equation}
     \delta_{u,v,k} = \sigma{(\epsilon_k)}\cdot{tanh(\hat{\delta}_{u, v, k})}.
\end{equation}
Here, $\sigma(\cdot)$ is a gated control unit, implemented by the classical sigmoid function, which controls how much information can be passed.

\subsubsection{Position Selector.} 
Knowledge transfer (or parameter sharing) has been proved to be potential for facilitating model learning. We take this inspiration to obtain the output $\mu_{u,v,k}$ of $k$-th position by summarize uplifts from subsequent positions (\ie $i \geq k$) as follows:
\begin{equation}
    \label{eq-out}
    \mu_{u, v, k} = \sum_{i = k}^K{\delta_{u, v, i}}.
\end{equation}
Clearly, the uplift for the $k$-th position $\delta_{u, v, k}$ is involved in the samples with position $i (i\leq{k})$, where the learned information can be shared among these samples.
To simplify the formulation and speed up the numerical computation, we construct a position matrix, which is denoted as $\mathbf{S}_{k}=[0 _{(0)}, \cdots, 0_{(k-1)}, 1_{(k)}, \cdots, 1_{(K)}]$ for position $k$. Then, we rewrite Eq.~\ref{eq-out} as follows:
\begin{equation}
    \label{eq-out_new}
    \mu_{u, v, k} = \mathbf{S}_{k}\cdot \bm{\Delta}_{u,v}^{T},
\end{equation}
where $\bm{\Delta}_{u,v}=[\delta_{u,v,0}, \delta_{u,v,1}, \cdots, \delta_{u,v,K}]$ is the position uplift vector. At last, given a user $u$ and an item $v$ with the position $k$, we estimate the click likelihood as follows:
\begin{equation}
    \hat{r}_{u,v,k}=\sigma(\mu_{u,v,k}).
\end{equation}

\begin{remark}
In general, the displayed position of item is a kind of posterior information, which cannot be obtained for online inference. Therefore, in practice, we only utilize the position data to perform position debiasing in training stage. And on the top of the well-trained model, we set all positions to 0 in default for online serving.
\end{remark}


\subsection{Model learning}
In {\model}, we intend to maximize the following posterior probability of model parameters $\Theta$ with observed interaction records $\mathcal{R} = \{u, v, k, t\}$ involving target user $u$ and item $v$ with corresponding position $k$ and click timestamp $t$.
\begin{equation}
    p(\Theta|\mathcal{D}) \propto p(\Theta) \cdot p(\mathcal{D}|\Theta),
\end{equation}
where $\Theta$ is the parameter set of {\model}. Here, the first $p(\Theta)$ measures the priori probability of model parameters $\Theta$, which can be regarded as the regularizer to avoid overfitting. Then, we mainly focus on the estimation of the second term $p(\mathcal{D}|\Theta)$, which can be factorized by minimizing its negative logarithm:
\begin{equation}
\begin{split}
    -\log p(\mathcal{D}|\Theta) 
          & = -\sum_{(u,v,k,t) \in \mathcal{R}}(\log p(v | k, u, \Theta) + \log p( t | u, v, k, \Theta)) \\
          & =  \sum_{(u,v,k,t) \in \mathcal{R}} \mathcal{C}(r_{u,v,k}, \hat{r}_{u,v,k})  \\
          & - \log{\lambda^*_v(t) + \sum_{v' \sim P_{neg}}{\int_0^T\lambda^*_{v'}(t)}dt} ).
\end{split}
\end{equation}
Here, $\mathcal{C}(\cdot, \cdot)$ is the cross entropy function and $P_{neg}$ is the noise distribution for the generator of negative samples.

%% file: 4-sec-sys.tex
\section{Online Deployment of {\model}}

\begin{figure}
    \centering
    \includegraphics[width=8cm]{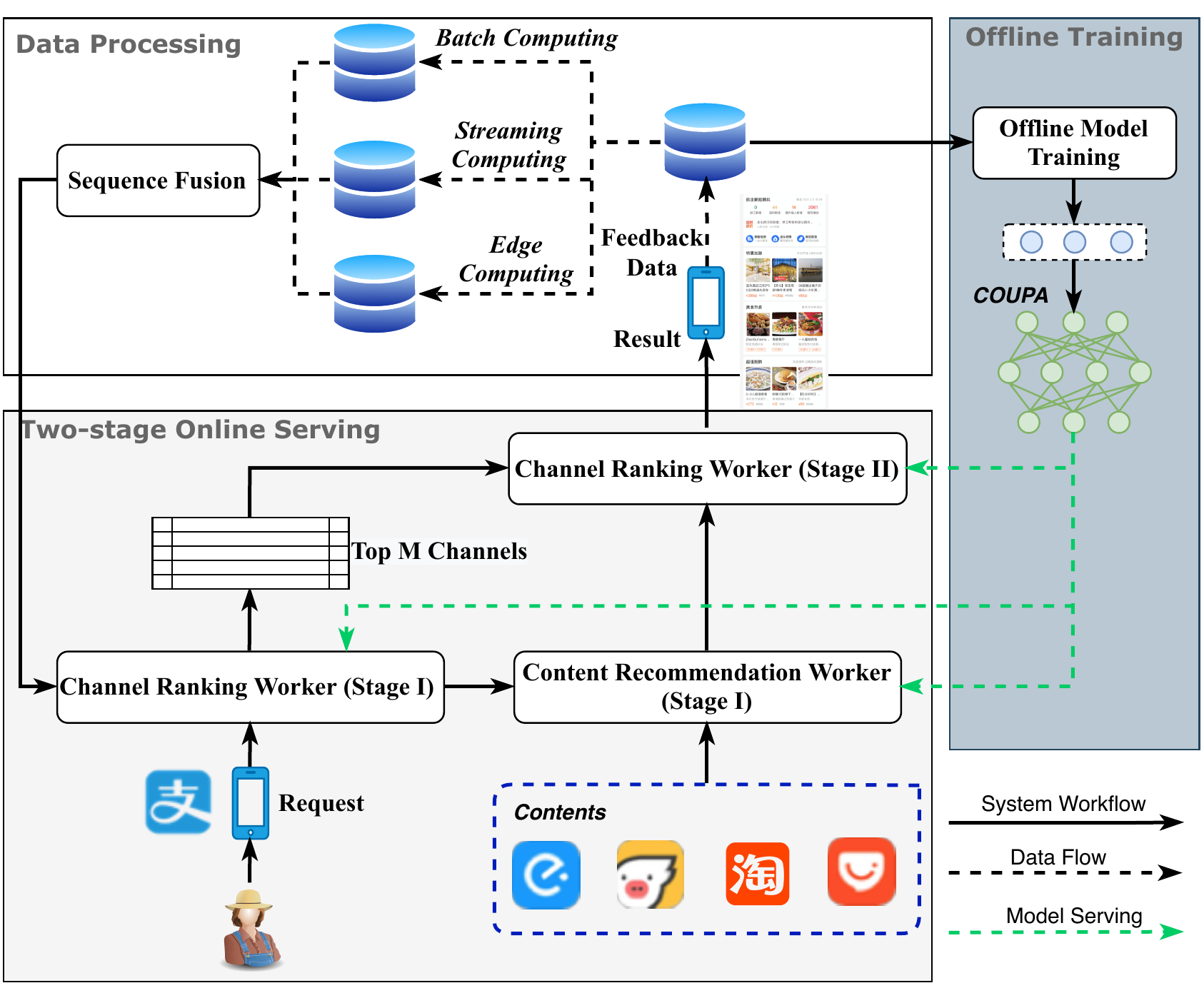}
    \caption{Online deployment of {\model} in Alipay.}
    \label{fig-deploy}
\end{figure}

In this section, we introduce the online deployment of the proposed {\model} in Alipay~\footnote{It is noteworthy that {\model} is also adapted for arbitrary recommendation scenarios in O2O service platforms.} to serve ranking tasks for channels and contents of each channel simultaneously, where the system design is still challenging. Most importantly, our temporal sequential model has urgent requirements of low delay for collecting real-time sequential data. Compared with traditional methods (\eg DeepFM), {\model} is relatively more complicated. It drives us to avoid amounts of delays for online inference, which needs to support high QPS (queries per second). To tackle these challenges, we carefully design the industrial system, roughly comprised of three main modules:  data processing, two-stage online serving and offline training.

\textbf{Data processing.}
User feedback data (\ie exposure and click) are necessary to 
train the recommender system, where the trade off between 
data storage and timely feedback need to be balanced. Hence, our principle is to
store 90-day click data in huge volume via cheap
storage that only guarantee latency in days, and timely feedback data in
tiny volume via edges that guarantee latency in seconds. As such,
we design the following batch, streaming and edge computing in a cooperative manner,
followed by a sequence fusion procedure.

\emph{Batch computing.} Based on the powerful MaxCompute~\footnote{https://www.alibabacloud.com/product/maxcompute}, an offline and low-cost computing platform developed by Alicloud~\footnote{https://www.alibabacloud.com/}, we perform batch computing for generating user click sequences in the last 90 days, involving hundreds of billions of interaction records, with a latency of days.

\emph{Streaming computing.} Streaming computing is mainly based on the frigate system (similar to Kafka~\footnote{https://kafka.apache.org/} and Storm~\footnote{https://storm.apache.org/} system) in the Ant Group, which directs at generating user click sequences in the last 2 days. Due to the affects of log back and real-time reporting, this part of the data has a latency of tens of seconds.

\emph{Edge computing.} 
Due to the urgent real-time demand of {\model}, we employ edge computing for collecting click sequences on the client in completely real time with a latency of only few seconds. Specifically, user click data will be captured by the client at once after reported, and recorded into local database. As soon as a user proposes a request, the client timely processes his/her click data in the last 3 hours and sends them to online servers.

\emph{Sequence fusion.} 
Sequence fusion is applied to fuse user click sequences in different time spans. It stitches the sequence data, generated by batch, streaming and edge computing, in chronological order and eliminates duplicates with timestamps. Subsequently, a complete sequence in the last 90 days is passed to {\model} as input.

\textbf{Two-stage online serving.}
This module is designed to relax delays of the online system for {\model} with two main stages. 
\emph{Stage I}: Channel Ranking Worker performs the coarse-grained ranking on channels with dozens of candidates based on {\model}, ignoring features of contents under channels. Then only top $M$ (\ie $M=5$) channels are selected to request their corresponding contents (\ie``Food Delive'' and ``Trave'' channels) recommender systems with tens of millions of candidates in Content Recommendation Worker, whose ranking model is {\model}.
\emph{Stage II}:  With contents under channels are determined in \emph{Stage I} as well as collected content features, Channel Ranking Worker performs the fine-grained ranking considering content features on top $M$ channels to decide their displayed orders.
In this way, system latency is greatly relaxed and ensures the stability of {\model} for online inference.


\textbf{Offline training.} 
This module aims at the training of the {\model} based on the user historical click records and online feature logs. With help of the MaxCompute and PAI~\footnote{https://www.alibabacloud.com/product/machine-learning} platform, we are capable of  training our model on tens of billions of data, whose details are stated in the Section~\ref{sec-model}.

%% file: 5-sec-exp.tex
\section{Experiments}

\begin{table}
\caption{Statistics of the datasets in offline experiments. \label{tab:off_data}}
\begin{tabular}{ccccccc}
\toprule
{} &{Food Delivery} & {Travel} & {Channel}\\
\midrule
{\# User}  & {1,388,754}& {1,323,131}& {8,396,541} \\
{\# Item} & {1,937,950}& {17,859}& {17} \\
{\# Train samples}      & {4,351,640}& {3,839,993}& {31,672,510} \\
{\# Validation samples}    & {200,000}& {200,000}& {200,000} \\
{\# Test samples}     & {3,385,818}& {3,186,217}& {1,682,579} \\
\bottomrule
\end{tabular}
\end{table}

In this section, we conduct a series of offline and online experiments to demonstrate the effectiveness of {\model}. 
In short, we aim at answering the following three research questions:
\begin{itemize}
    \item \textbf{RQ1}: Does our proposed model outperforms other state-of-the-art methods on both temporal recommendation and position debiasing task.
    \item \textbf{RQ2}: Does our proposed model achieve superior performance in real-world scenarios.
    \item \textbf{RQ3}: Does our proposed model intuitively provide convincing evidences for recommendation.
\end{itemize}

\subsection{Experimental Setup}

\subsubsection{Dataset} We collect three real-world datasets for offline evaluation from  Alipay, namely \textbf{Food Delivery} , \textbf{Travel} and \textbf{Channel} dataset. 
Specifically, we extract the interaction records from ``2021-01-07'' to ``2021-01-13'' for training and remain records on ``2021-01-13'' for test. Moreover, we leave out 200, 000 samples of training data as the validation set for parameter tuning. For each user in the extracted dataset, we collect his/her interaction records in the last three months to construct recent behaviors. We only preserve the users who have more than 10 behaviors and keep the ratio of positive and negative samples as 1 : 6. Also, abundant features are extract to effectively characterize users, items and associated scenarios, which are presented in the Section 4.
Statistics of the datasets are shown in Table~\ref{tab:off_data}.


\subsubsection{Baselines}
In our experiments, we compare our approach with several state-of-the-art baselines, developed for sequential recommendation (\ie \textbf{DeepFM~\cite{guo2017deepfm}}, \textbf{GRU4Rec~\cite{hidasi2015session}}, \textbf{DIN~\cite{zhou2018deep}}, \textbf{DIEN~\cite{zhou2019deep}}, \textbf{SASRec~\cite{kang2018self}}, \textbf{Time-LSTM~\cite{zhu2017next}}) and position debiasing (\ie \textbf{YoutubeRank~\cite{zhao2019recommending}} and \textbf{PAL~\cite{guo2019pal}}), respectively. Considering the industrial settings, the selected baselines have potential for scaling up to huge volume of datasets.
\begin{itemize}
    \item \textbf{DeepFM~\cite{guo2017deepfm}} : It is a typical feature based recommendation method consisting of a factorization machine (FM) component and a deep neural network (DNN) component.  
    \item \textbf{GRU4Rec~\cite{hidasi2015session}} : It is a sequential recommendation method with a RNN structured GRU part for user behaviors modelling.
    \item \textbf{DIN~\cite{zhou2018deep}} : It is a sequential recommendation method with an attention mechanism to exploit related user behaviors. 
    \item \textbf{DIEN~\cite{zhou2019deep}} : It is an improved version of DIN by considering the interest evolving process through GRU with an attention update gate.
    \item \textbf{SASRec~\cite{kang2018self}} : It is a representative sequential recommendation method using a left-to-right Transformer module to capture user behaviors. 
    \item \textbf{Time-LSTM~\cite{zhu2017next}} : It is a temporal recommendation method, equipping a LSTM structure with time gates for time interval modelling.
    
    \item \textbf{YoutubeRank~\cite{zhao2019recommending}}: It is a industrial video ranking system for Youtube, which employs the Multi-gate Mixture-of-Experts to optimize multiple ranking objectives and adopt a Wide\&Deep framework for mitigating the selection bias.
    
    \item \textbf{PAL~\cite{guo2019pal}}: It is a position bias aware learning framework for CTR prediction, which jointly and simultaneously optimizes the probability that a user see and click  a target item.
\end{itemize}

Moreover, {\model} has two variant:
\begin{itemize}
    \item \textbf{{\model}}$_{\mathcal{T}}$: It is a variant of {\model}, which only models user's temporal preference over continuous time
    \item \textbf{{\model}}$_{\mathcal{P}}$: It is another variant of {\model}, which  only aims at position debiasing task. 
\end{itemize}

\subsubsection{Implementation Details \label{app-impl}}
We implement all methods on parameter server based distributed learning systems~\cite{zhou2017kunpeng} with Tensorflow 1.13. All parameters for baselines are optimized in the validation set as mentioned above, and we briefly present the optimal parameters in our following experiments for reproducibility. For all methods, we perform Adam for optimization with learning rate 1e-4. We set embedding size of each feature as 8 and set the architecture of MLP as [256, 128, 64]. For sequential methods, the max length of sequence is set as 50. For SASRec, two self-attention blocks are used. We run each method ten times and average the results as the final performance.

\subsubsection{Evaluation Metrics}
In our experiments, we employ several widely used metrics to evaluate the offline and online performance of all approaches, respectively. To evaluate the offline performance (\ie Section~\ref{sec-exp-tr}), we adopt the group weighted area under curve (\textbf{GAUC}), which is a more fine-grained metric in industrial settings since it is more relevant to online performance. 
Formally, we can calculate the GAUC as follows:
\begin{equation}
\text{GAUC} = \frac{\sum_g{w_g \times \text{AUC}_g}}{\sum_g{w_g}}, 
\end{equation}
where $w_g$ is the weight of group $g$ (\ie the number of samples in group $g$) and AUC$_g$ is the AUC for group $g$.
Moreover, we also report the relative improvement (\textbf{RI}) \wrt GAUC of our approach over compared baselines, which is defined as:
\begin{equation}
    \text{RI} = \frac{|\text{GAUC}_{ours} - \text{GAUC}_{base}|}{\text{GAUC}_{base}} \times 100\%,
\end{equation}
where $|\cdot|$ is the absolute value, $GAUC_{ours/base}$ refers to the performance of our approach and corresponding baselines, respectively.
Note that it is remarkable in our industrial scenarios (especially the channel rank scenario) that only \textbf{0.1\%} improvement \wrt GAUC is achieved. To evaluate the online performance (\ie Section~\ref{sec-exp-pd} and~\ref{sec-exp-online}), we use the metric \textbf{IPV} and \textbf{CTR}, which are commonly adopted in various industrial online systems. 

\begin{table}

\caption{Performance comparison on three offline datasets. We use the \textbf{bold font} to mark the best performance for each comparison. We use ``** (or *)''  to indicate that the improvement of {\model}$_{\mathcal{T}}$ over the best baseline is significant base on the $t$-test at the significance level of 0.01 (0.05). \label{tab:perf_off}}
\setlength{\tabcolsep}{0.8mm}{
\begin{tabular}{ccccccc}
\toprule
{} & \multicolumn{2}{c}{Food Delivery} & \multicolumn{2}{c}{Travel} & \multicolumn{2}{c}{Channel}\\
{} & {GAUC}& {RI}& {GAUC}& {RI}& {GAUC}& {RI} \\
\midrule
{DeepFM}      & {0.7534} &{+3.31\%}& {0.7198} &{+4.86\%}& {0.8837} &{+0.26\%} \\
{GRU4Rec}    & {0.7682} &{+1.32\%}& {0.7373} &{+2.37\%}& {0.8839} &{+0.15\%} \\
{DIN}     & {0.7682} &{+1.32\%}& {0.7371} &{+2.40\%}& {0.8846}  &{+0.16\%}\\
{DIEN}  & {0.7701} &{+1.08\%}& {0.7421} &{+1.71\%}  &{0.8849}& {+0.12\%}\\
{SASRec} & {0.7688} &{+1.25\%}& {0.7362} &{+2.52\%} &{0.8843}& {+0.19\%} \\
{Time-LSTM}  & {0.7698} &{+1.12\%}& {0.7399} &{+2.01\%} &{0.8842}& {+0.20\%} \\
\midrule
{\model$_{\mathcal{T}}$}          & {\textbf{0.7784**}}&{-}& {\textbf{0.7548**}}&{-}& {\textbf{0.8860*}} &{-}\\
\bottomrule
\end{tabular}}
\end{table}



\begin{table}
\caption{Performance comparison for position debiasing task on three online scenarios. We report the improvement ratio over the baseline YoutubeRank.\label{tab:perf_pb}}
\setlength{\tabcolsep}{2.0mm}{
\begin{tabular}{ccccc}
\toprule
{} &{} & {CTR}& {IPV} \\
\midrule
{} &{YoutubeRank}  & {-} & {-}   \\
{Food Delivery} &{PAL}  & {+1.07\%} & {+1.03\%}   \\
{} &{\model$_{\mathcal{P}}$}& {\textbf{+1.44\%}} & {\textbf{+1.28\%}} \\
\bottomrule
{} &{YoutubeRank}  & {-} & {-}  \\
{Travel} &{PAL}  & {+0.78\%} & {+0.84\%}   \\
{} &{\model$_{\mathcal{P}}$} & {\textbf{+1.22}\%}& {\textbf{+1.17\%}} \\
\bottomrule
{} &{YoutubeRank}  & {-} & {-}   \\
{Channel} &{PAL}  & {+1.26\%} & {+1.50\%}   \\
{} &{\model$_{\mathcal{P}}$}& {\textbf{+2.04\%}} & {\textbf{+2.18\%}} \\
\bottomrule
\end{tabular}}
\end{table}

\begin{figure*}[t]
\centering
\subfigure[Food delivery]{
\begin{minipage}[t]{0.32\linewidth}
\centering
\includegraphics[width=6cm]{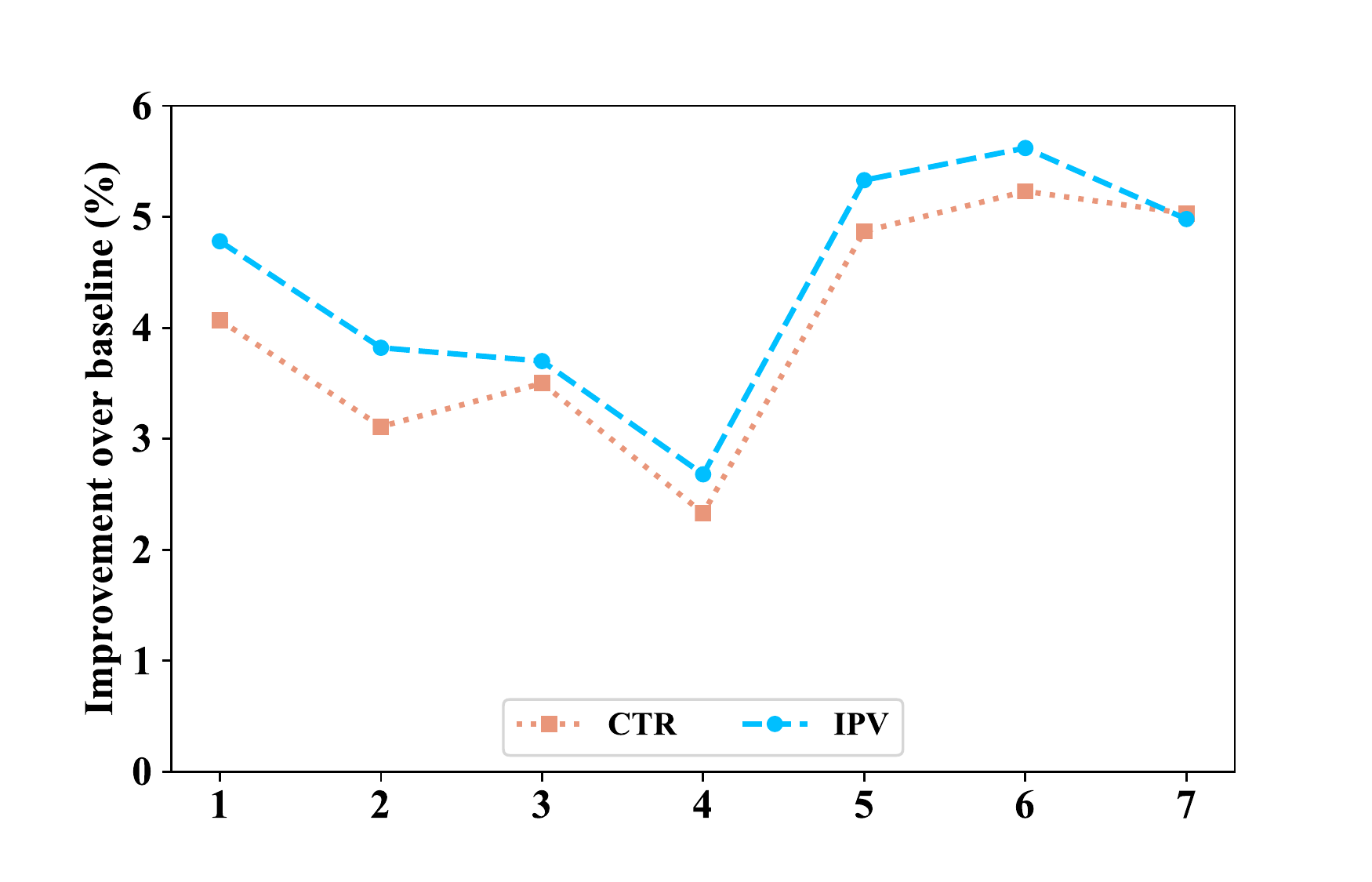}
\end{minipage}
}
\subfigure[Travel]{
\begin{minipage}[t]{0.32\linewidth}
\centering
\includegraphics[width=6cm]{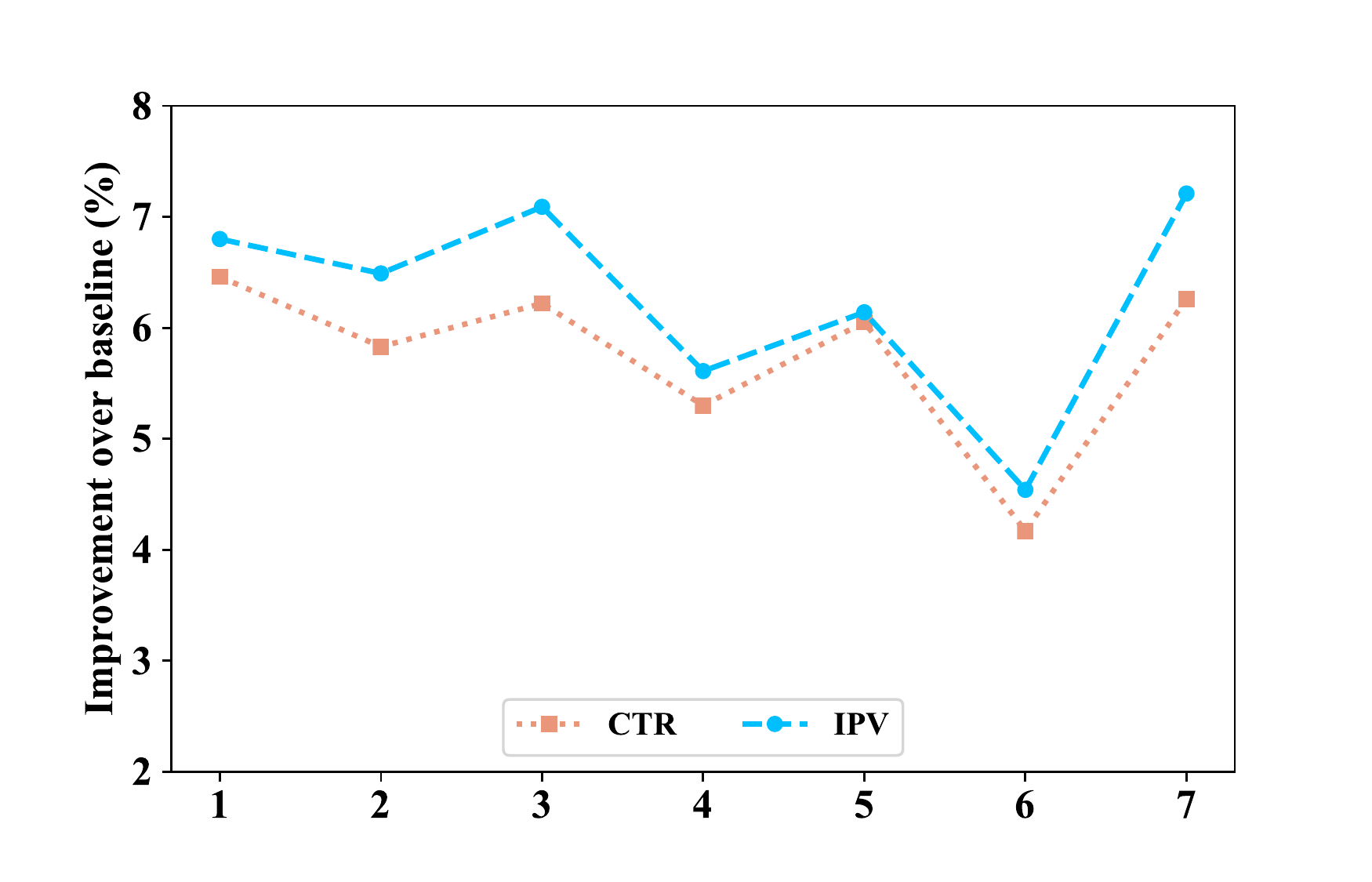}
\end{minipage}
}
\subfigure[Channel]{
\begin{minipage}[t]{0.32\linewidth}
\centering
\includegraphics[width=6cm]{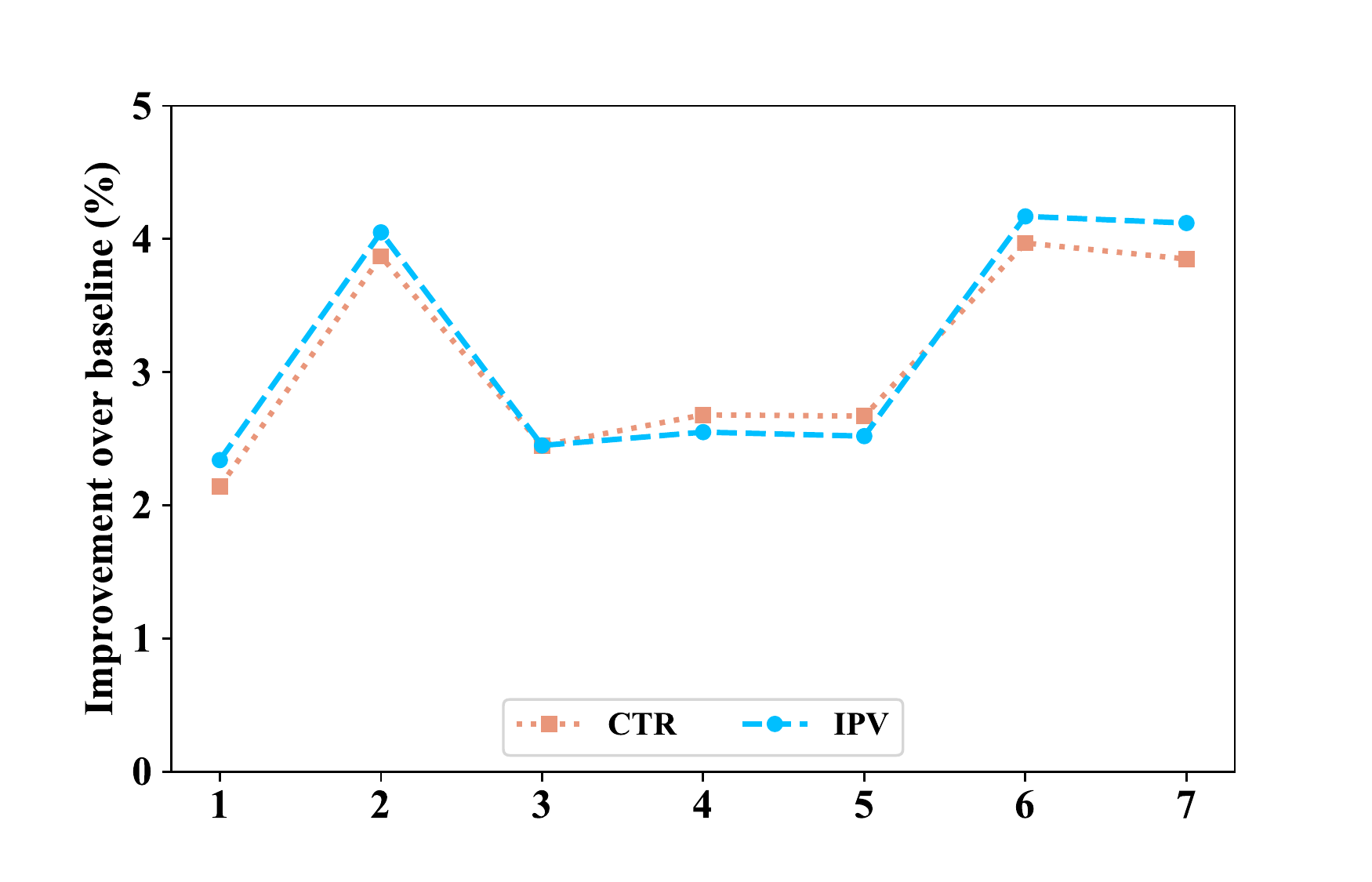}
\end{minipage}
}
\caption{Online performance on three real scenarios in Alipay. $y$-axis denotes the improvement ratio over the best baseline.\label{fig-online_performance}}
\end{figure*}

\subsection{Performance Comparison (\textbf{RQ1})}
\subsubsection{Temporal Recommendation \label{sec-exp-tr}}
We evaluate the performance of temporal recommendation on three offline datasets. Since the performance of position debiasing cannot be evaluated offline, we only show the performance of the variant {\model}$_{\mathcal{T}}$ and select DeepFM, GRU4Rec, DIN, DIEN, SASRec and Time-LSTM as baselines for comparison. 

We report the comparison results on three offline datasets in Table~\ref{tab:perf_off} and summarize the major findings as follows.
(i) {\model}$_{\mathcal{T}}$ is consistently better than all the baselines with statistical significance, demonstrating its effectiveness to learn time aware preference for recommendation. Note that the improvement on the Channel dataset is relatively weak since there are extremely few items in this dataset. Nevertheless,  such a slight gain still is remarkable for online performance, which will be shown in Section~\ref{sec-exp-online}.
(ii) Among baselines, the performance improvement of GRU4Rec, DIN and SASRec \wrt DeepFM reveals effectiveness of user historical behaviors for inferring his/her preferences, while the performance improvements of DIEN \wrt DIN and Time-LSTM \wrt GRU4Rec imply the importance of modelling interest evolution for users over time.
(iii) {\model}$_{\mathcal{T}}$ still significantly outperforms both state-of-the-art sequential and temporal recommendation methods, attributed to the excellent ability for summarizing influence from historical interactions for recommendation through continuous time aware attention mechanism and point process.
Overall, {\model}$_{\mathcal{T}}$ achieves performance improvement over the best baseline by 1.08\%, 1.71\% and 0.12\% on the three offline datasets, respectively.

\subsubsection{Position Debiasing \label{sec-exp-pd}}
We examine the performance of position debiasing on three online scenarios (\ie Food delivery, Travel and Channels). For fair comparison, we only report the performance of {\model}$_{\mathcal{P}}$ and select YoutubeRank and PAL as baselines for comparison. For convenience,  we report the relative improvement ratios \wrt YoutubeRank. We perform the evaluation with 7 days and report average results for each method in Table~\ref{tab:perf_pb}.

From the results, we observe that {\model}$_{\mathcal{P}}$ achieves the best performance with statistical significance in all scenarios in term of both IPV and CTR metrics,
clearly demonstrating the superiorities of {\model}$_{\mathcal{P}}$ for performing position debiasing in a personalized manner. It is noteworthy that {\model}$_{\mathcal{P}}$ works extremely well on the scenerio of channel rank. An intuitive explanation is that the position bias issue of this scenario is more serious since each channel occupies a large area and  the corresponding click distributions are indeed very skew, which has been analyzed in Section 2. Obviously, these issues are unable to be well handled by the YoutubeRank and PAL.

\subsection{Online Performance (\textbf{RQ2}) \label{sec-exp-online}}
To further verify the effectiveness of the proposed {\model} in the real-world settings, we conduct a series of experiments for online services. Similarly, we deploy {\model} into three scenarios (\ie Food delivery, Travel and Channels) in Alipay,  comparing it with the existed deployed baseline in our real system.
Specifically, for each scenario, we conduct a bucket testing (\ie A/B testing) online to evaluate the users’ responses to {\model} against the baseline, where we select one bucket for {\model} and another for the baseline. We perform the evaluation from ``2020-01-13'' to ``2020-01-20'' and present performance comparison results in Figure~\ref{fig-online_performance}. For convenience, we report the relative improvement ratios \wrt the selected baseline. 

From the results, We observe that, compared to the best baseline used in our real system, {\model} consistently and significantly yields performance improvement by a large margin in three scenarios across all online metrics, which further demonstrates superior abilities of {\model} for capturing user preferences over time as well as mitigating the position bias in real-world applications. Overall, for the three real scenarios, {\model} gains the average improvement of 4.06\%, 6.37\%, 3.11\% for IPV, 3.73\%, 5.63\%, 2.47\% for CTR.

\begin{figure}[h]
	\centering
	\subfigure[] {
		\includegraphics[width=40mm,height=30mm]{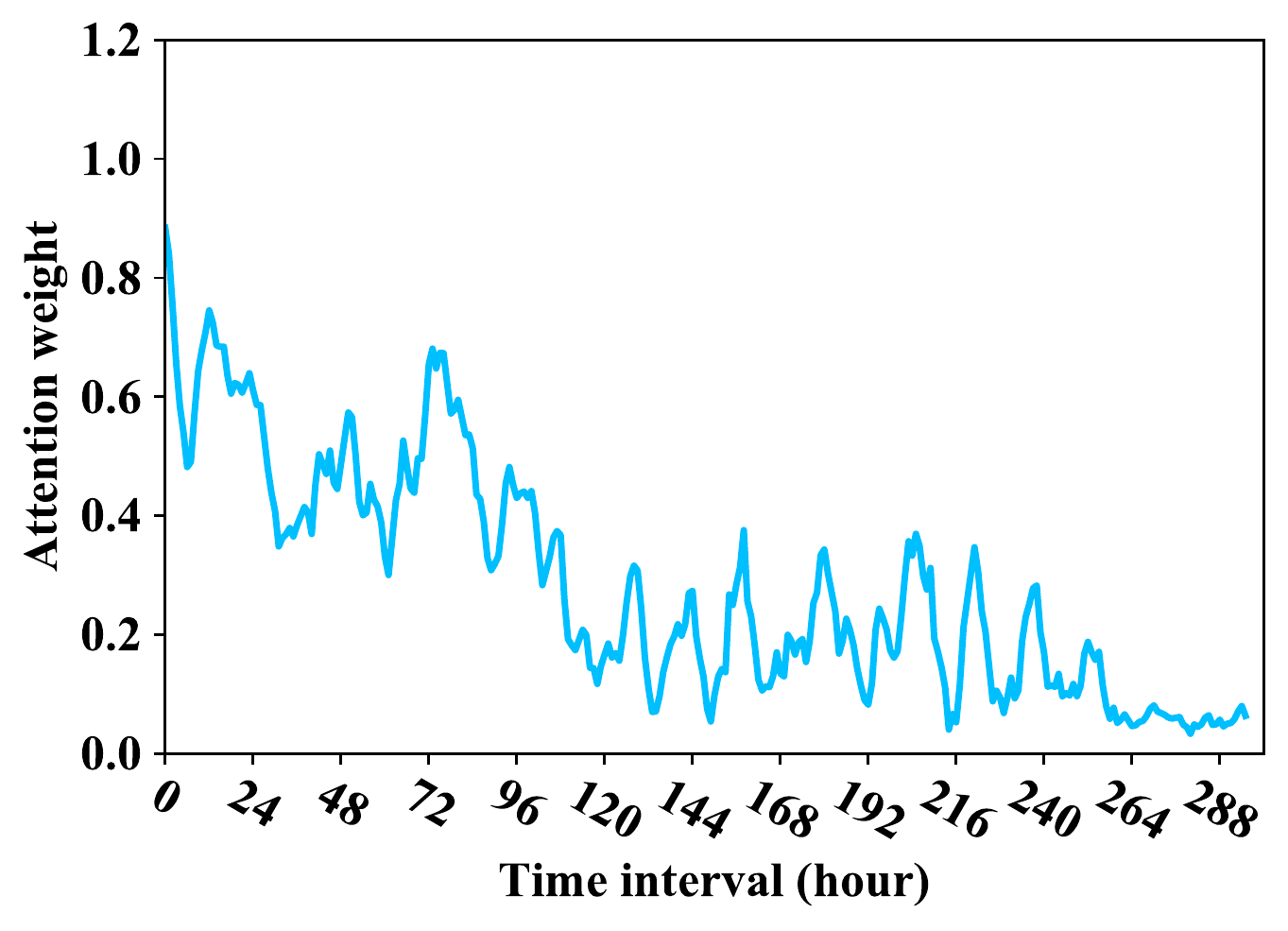}
	}
	\subfigure[] {
		\includegraphics[width=40mm,height=30mm]{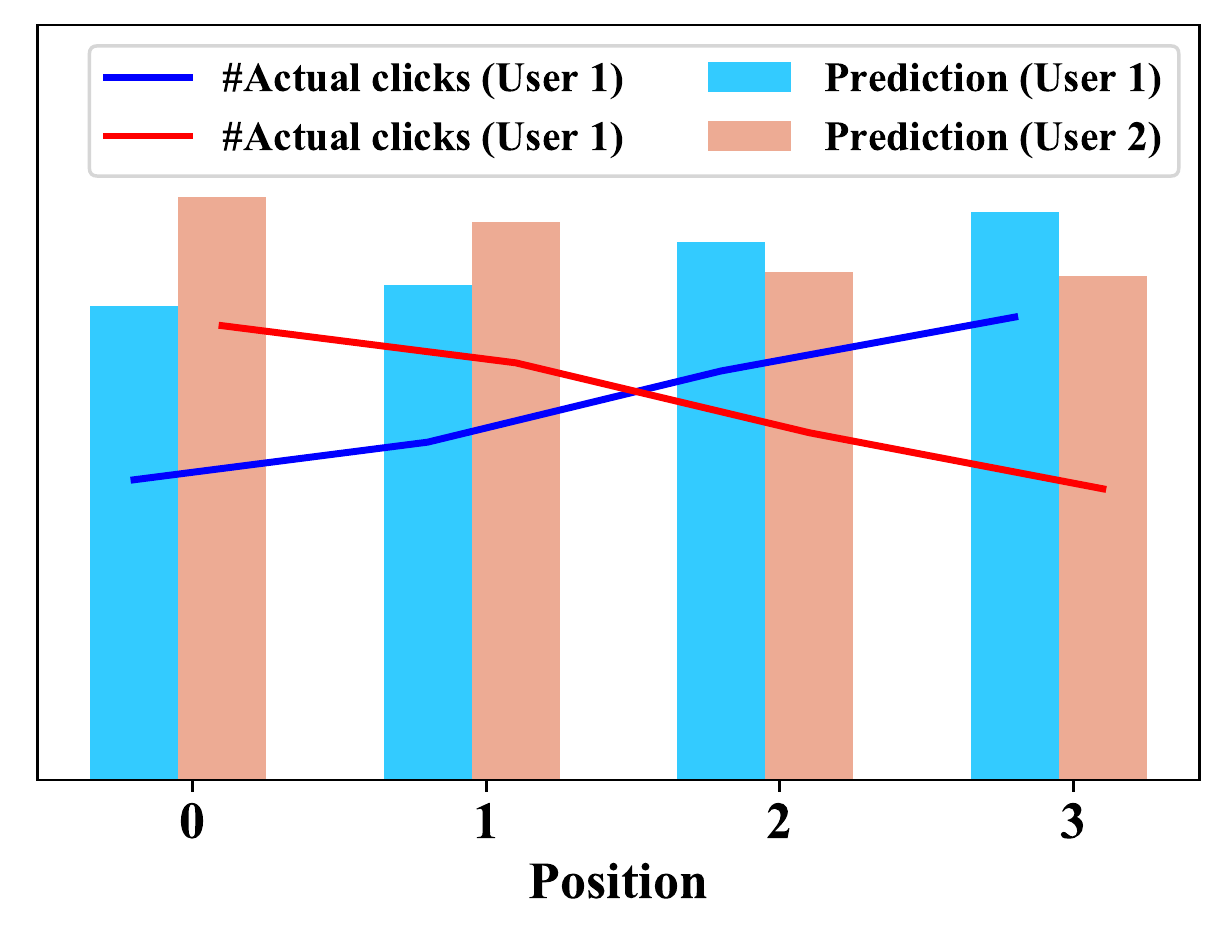}
	}
	\caption{Case studies for time (a) and position bias (b) aware preference modelling, respectively. The smaller the number of $x$-axis in (b), the higher the position.}
	\label{fig:case}
\end{figure}

\subsection{Case Study (\textbf{RQ3})}
At last, we conduct case studies to show how {\model} intuitively provides reasonable evidences for recommendation in term of continuous time and position bias, respectively.

From the temporal perspective, we select a certain user from dataset of Food delivery channel and plot his/her attention weights \wrt time interval of historical behaviors in Figure~\ref{fig:case} (a). We have following two observations: (i) With the increase of time interval, the corresponding attention weight decreases, demonstrating user's current preference is extremely affected by recent behaviors. (ii) The change of the attention weight has periodical trend, which is consistent with our findings in Section 2. In sum, both observations intuitively verifies that {\model} is capable of summarizing influence from historical interactions for recommendation.

From the perspective of position bias, we select two typical user and plot the average prediction score for each position $k$(\ie $\sum_{v}{\mu_{u,v,k}}$ according to Eq.~\ref{eq-out}) as well as their actual click trends in Figure~\ref{fig:case} (b). We only present the performance for the top 4 positions (\ie ``0'' $\rightarrow$ ``3'') since they are most impressed to some extent by users in the channel rank scenario of Alipay. As the position changed, we find the distinct trends of average prediction score for the two selected users (\ie increase for ``User 1'' and decrease for ``User 2''). Not surprisingly, their score distributions predicted by {\model} are consistent with their trends of actual clicks  \wrt each position. Above analyses further verify the strong capability of {\model} for capturing the personalized bias based on the real distribution over positions.


%% file: 6-sec-rel.tex
\section{Related Work}

Recent years have witnessed the great success of sequential recommendation methods, which devote to predicting user interests based on his/her historical behavior sequences. Early works mainly focus on modelling behavior sequences with Markov chains~\cite{rendle2010factorizing,he2016fusing,zimdars2013using}. Due to the ability of capturing complex feature interaction, the deep neural network (DNN) has yielded excellent performance in various applications. To marry up the advantage of DNN and sequential recommendation, a series of methods attempt to extract user preferences from his/her previous behaviors through recursive neural network (RNN)~\cite{hidasi2015session,tan2016improved}, convolutional neural network (CNN)~\cite{xu2019recurrent,tang2018personalized,yan2019cosrec} and memory network~\cite{chen2018sequential} based structures. Meanwhile, recently emerging neural attention mechanism~\cite{vaswani2017attention} is commonly adopted in  characterizing long- and short-term preferences in a more fine-grained principle~\cite{zhou2018deep,zhou2019deep,kang2018self,luocollaborative}. Furthermore, many recent studies have proved that temporal information is of crucial importance for recommendation, which implies users' underlying reasons to click/purchase a target item. Therefore, a few works are proposed to incorporate such temporal information into sequential recommendation with extended LSTM structure~\cite{zhu2017next}, temporal point process~\cite{bai2019ctrec} and time aware neural network~\cite{ye2020time}. Here, we please refer the reader to recent surveys~\cite{fang2020deep,wu2021survey} for a thorough review,. Unfortunately, due to the time sensitivity of user preferences, how to fully exploit influences derived from historical interaction records still remains challenging. Moreover, in the industrial settings, automatically collecting real-time sequential data and guaranteeing low delays for on online serving also need to be carefully considered.

On the other hand, recent years have seen a surge of efforts on exploring the impacts of the biases and performing necessary debiasing in recommender systems. In our paper, we center on the widely studied position bias, which denotes that users are more likely to click the higher ranked items regardless of the corresponding relevance. To solve the position bias, a number of works have been proposed, which roughly fall into two lines. The first line makes some basic hypotheses about the user browsing behaviors. Specifically, these methods maximize the likelihood of the observed interaction data for the true relevance feedback with self-defined examination hypothesises~\cite{craswell2008experimental,dupret2008user,chapelle2009dynamic},  personalized transition probability based cascade models~\cite{guo2009click,zhu2010novel} and state-of-the-art deep recurrent survival model~\cite{jin2020deep}. The other line is based on the recently emerging counterfactual learning~\cite{joachims2017unbiased,wang2018position}, which weights each sample with a position aware value and collects the user feedback with inverse propensity score~\cite{agarwal2019general}. Following this line, some recent works~\cite{qin2020attribute,ai2018unbiased} further adopt the dual leaning for jointly learning a propensity model as well as a recommendation model with well-designed EM algorithms. To learn more works about debising in recommender systems, please refer to the elaborate survey~\cite{chen2020bias}.
Nevertheless, these methods mainly assume the independence of positions, while the personalization of position bias is still unexplored.

%% file: 7-sec-con.tex
\section{Conclusion and Future Work}
In this paper, we proposed a novel continuous time and position aware recommender system for O2O service platforms (\eg Alipay), called {\model}, which comprehensively takes user preferences towards temporal patterns and position biases into consideration. 
To better learn temporal patterns in real-world applications, we develop a continuous time aware point process equipped with continuous time aware attention mechanism to chronologically summarize influences derived from historical behaviors for recommendation.
Moreover, a position selector module cooperated with a Multi-gate Mixture-of-Experts (MMoE) block and  a Gated Linear Units (GLU) black is introduced for mitigating position bias in a personalized manner. 
At last, we devoted to the design and implementation of our whole system, which jointly employs edge, streaming and batch computing for capturing the real-time user behaviors and adopt a two-stage mode for efficient online serving. Extensive offline and online experiments demonstrated the superiority of our proposed {\model}.

Currently, our approach is able to effectively extract user preferences from historical interaction records for recommendation. While, there exist a number of users/items with sparse interactions in real-world scenarios, for which current approach fails to learn high-quality representations. As future work, we will consider how to incorporate graph structure data (\eg social network and knowledge graph) for compensating these cold-start users/items. In addition, we will also consider adapting our approach to well deal with geographical information, which plays an important role in O2O service platforms.

%% file: main.bbl

\begin{thebibliography}{47}


\ifx \showCODEN    \undefined \def \showCODEN     #1{\unskip}     \fi
\ifx \showDOI      \undefined \def \showDOI       #1{#1}\fi
\ifx \showISBNx    \undefined \def \showISBNx     #1{\unskip}     \fi
\ifx \showISBNxiii \undefined \def \showISBNxiii  #1{\unskip}     \fi
\ifx \showISSN     \undefined \def \showISSN      #1{\unskip}     \fi
\ifx \showLCCN     \undefined \def \showLCCN      #1{\unskip}     \fi
\ifx \shownote     \undefined \def \shownote      #1{#1}          \fi
\ifx \showarticletitle \undefined \def \showarticletitle #1{#1}   \fi
\ifx \showURL      \undefined \def \showURL       {\relax}        \fi
\providecommand\bibfield[2]{#2}
\providecommand\bibinfo[2]{#2}
\providecommand\natexlab[1]{#1}
\providecommand\showeprint[2][]{arXiv:#2}

\bibitem[\protect\citeauthoryear{Agarwal, Takatsu, Zaitsev, and
  Joachims}{Agarwal et~al\mbox{.}}{2019}]%
        {agarwal2019general}
\bibfield{author}{\bibinfo{person}{Aman Agarwal}, \bibinfo{person}{Kenta
  Takatsu}, \bibinfo{person}{Ivan Zaitsev}, {and} \bibinfo{person}{Thorsten
  Joachims}.} \bibinfo{year}{2019}\natexlab{}.
\newblock \showarticletitle{A general framework for counterfactual
  learning-to-rank}. In \bibinfo{booktitle}{\emph{SIGIR}}.
  \bibinfo{pages}{5--14}.
\newblock


\bibitem[\protect\citeauthoryear{Ai, Bi, Luo, Guo, and Croft}{Ai
  et~al\mbox{.}}{2018}]%
        {ai2018unbiased}
\bibfield{author}{\bibinfo{person}{Qingyao Ai}, \bibinfo{person}{Keping Bi},
  \bibinfo{person}{Cheng Luo}, \bibinfo{person}{Jiafeng Guo}, {and}
  \bibinfo{person}{W~Bruce Croft}.} \bibinfo{year}{2018}\natexlab{}.
\newblock \showarticletitle{Unbiased learning to rank with unbiased propensity
  estimation}. In \bibinfo{booktitle}{\emph{SIGIR}}. \bibinfo{pages}{385--394}.
\newblock


\bibitem[\protect\citeauthoryear{Bai, Zou, Zhao, Du, Liu, Nie, and Wen}{Bai
  et~al\mbox{.}}{2019}]%
        {bai2019ctrec}
\bibfield{author}{\bibinfo{person}{Ting Bai}, \bibinfo{person}{Lixin Zou},
  \bibinfo{person}{Wayne~Xin Zhao}, \bibinfo{person}{Pan Du},
  \bibinfo{person}{Weidong Liu}, \bibinfo{person}{Jian-Yun Nie}, {and}
  \bibinfo{person}{Ji-Rong Wen}.} \bibinfo{year}{2019}\natexlab{}.
\newblock \showarticletitle{Ctrec: A long-short demands evolution model for
  continuous-time recommendation}. In \bibinfo{booktitle}{\emph{SIGIR}}.
  \bibinfo{pages}{675--684}.
\newblock


\bibitem[\protect\citeauthoryear{Chapelle and Zhang}{Chapelle and
  Zhang}{2009}]%
        {chapelle2009dynamic}
\bibfield{author}{\bibinfo{person}{Olivier Chapelle} {and} \bibinfo{person}{Ya
  Zhang}.} \bibinfo{year}{2009}\natexlab{}.
\newblock \showarticletitle{A dynamic bayesian network click model for web
  search ranking}. In \bibinfo{booktitle}{\emph{WWW}}. \bibinfo{pages}{1--10}.
\newblock


\bibitem[\protect\citeauthoryear{Chen, Dong, Wang, Feng, Wang, and He}{Chen
  et~al\mbox{.}}{2020}]%
        {chen2020bias}
\bibfield{author}{\bibinfo{person}{Jiawei Chen}, \bibinfo{person}{Hande Dong},
  \bibinfo{person}{Xiang Wang}, \bibinfo{person}{Fuli Feng},
  \bibinfo{person}{Meng Wang}, {and} \bibinfo{person}{Xiangnan He}.}
  \bibinfo{year}{2020}\natexlab{}.
\newblock \showarticletitle{Bias and Debias in Recommender System: A Survey and
  Future Directions}.
\newblock \bibinfo{journal}{\emph{arXiv preprint arXiv:2010.03240}}
  (\bibinfo{year}{2020}).
\newblock


\bibitem[\protect\citeauthoryear{Chen, Xu, Zhang, Tang, Cao, Qin, and Zha}{Chen
  et~al\mbox{.}}{2018}]%
        {chen2018sequential}
\bibfield{author}{\bibinfo{person}{Xu Chen}, \bibinfo{person}{Hongteng Xu},
  \bibinfo{person}{Yongfeng Zhang}, \bibinfo{person}{Jiaxi Tang},
  \bibinfo{person}{Yixin Cao}, \bibinfo{person}{Zheng Qin}, {and}
  \bibinfo{person}{Hongyuan Zha}.} \bibinfo{year}{2018}\natexlab{}.
\newblock \showarticletitle{Sequential recommendation with user memory
  networks}. In \bibinfo{booktitle}{\emph{WSDM}}. \bibinfo{pages}{108--116}.
\newblock


\bibitem[\protect\citeauthoryear{Cheng, Koc, Harmsen, Shaked, Chandra, Aradhye,
  Anderson, Corrado, Chai, Ispir, et~al\mbox{.}}{Cheng et~al\mbox{.}}{2016}]%
        {cheng2016wide}
\bibfield{author}{\bibinfo{person}{Heng-Tze Cheng}, \bibinfo{person}{Levent
  Koc}, \bibinfo{person}{Jeremiah Harmsen}, \bibinfo{person}{Tal Shaked},
  \bibinfo{person}{Tushar Chandra}, \bibinfo{person}{Hrishi Aradhye},
  \bibinfo{person}{Glen Anderson}, \bibinfo{person}{Greg Corrado},
  \bibinfo{person}{Wei Chai}, \bibinfo{person}{Mustafa Ispir}, {et~al\mbox{.}}}
  \bibinfo{year}{2016}\natexlab{}.
\newblock \showarticletitle{Wide \& deep learning for recommender systems}. In
  \bibinfo{booktitle}{\emph{RecSys Workshop}}. \bibinfo{pages}{7--10}.
\newblock


\bibitem[\protect\citeauthoryear{Covington, Adams, and Sargin}{Covington
  et~al\mbox{.}}{2016}]%
        {covington2016deep}
\bibfield{author}{\bibinfo{person}{Paul Covington}, \bibinfo{person}{Jay
  Adams}, {and} \bibinfo{person}{Emre Sargin}.}
  \bibinfo{year}{2016}\natexlab{}.
\newblock \showarticletitle{Deep neural networks for youtube recommendations}.
  In \bibinfo{booktitle}{\emph{RecSys}}. \bibinfo{pages}{191--198}.
\newblock


\bibitem[\protect\citeauthoryear{Craswell, Zoeter, Taylor, and Ramsey}{Craswell
  et~al\mbox{.}}{2008}]%
        {craswell2008experimental}
\bibfield{author}{\bibinfo{person}{Nick Craswell}, \bibinfo{person}{Onno
  Zoeter}, \bibinfo{person}{Michael Taylor}, {and} \bibinfo{person}{Bill
  Ramsey}.} \bibinfo{year}{2008}\natexlab{}.
\newblock \showarticletitle{An experimental comparison of click position-bias
  models}. In \bibinfo{booktitle}{\emph{WSDM}}. \bibinfo{pages}{87--94}.
\newblock


\bibitem[\protect\citeauthoryear{Dauphin, Fan, Auli, and Grangier}{Dauphin
  et~al\mbox{.}}{2017}]%
        {dauphin2017language}
\bibfield{author}{\bibinfo{person}{Yann~N Dauphin}, \bibinfo{person}{Angela
  Fan}, \bibinfo{person}{Michael Auli}, {and} \bibinfo{person}{David
  Grangier}.} \bibinfo{year}{2017}\natexlab{}.
\newblock \showarticletitle{Language modeling with gated convolutional
  networks}. In \bibinfo{booktitle}{\emph{ICML}}. \bibinfo{pages}{933--941}.
\newblock


\bibitem[\protect\citeauthoryear{Dupret and Piwowarski}{Dupret and
  Piwowarski}{2008}]%
        {dupret2008user}
\bibfield{author}{\bibinfo{person}{Georges~E Dupret} {and}
  \bibinfo{person}{Benjamin Piwowarski}.} \bibinfo{year}{2008}\natexlab{}.
\newblock \showarticletitle{A user browsing model to predict search engine
  click data from past observations.}. In \bibinfo{booktitle}{\emph{SIGIR}}.
  \bibinfo{pages}{331--338}.
\newblock


\bibitem[\protect\citeauthoryear{Fang, Zhang, Shu, and Guo}{Fang
  et~al\mbox{.}}{2020}]%
        {fang2020deep}
\bibfield{author}{\bibinfo{person}{Hui Fang}, \bibinfo{person}{Danning Zhang},
  \bibinfo{person}{Yiheng Shu}, {and} \bibinfo{person}{Guibing Guo}.}
  \bibinfo{year}{2020}\natexlab{}.
\newblock \showarticletitle{Deep Learning for Sequential Recommendation:
  Algorithms, Influential Factors, and Evaluations}.
\newblock \bibinfo{journal}{\emph{ACM Transactions on Information Systems}}
  \bibinfo{volume}{39}, \bibinfo{number}{1} (\bibinfo{year}{2020}),
  \bibinfo{pages}{1--42}.
\newblock


\bibitem[\protect\citeauthoryear{Gong, Jiang, Feng, Hu, Zhao, Liu, and Ou}{Gong
  et~al\mbox{.}}{2020}]%
        {gong2020edgerec}
\bibfield{author}{\bibinfo{person}{Yu Gong}, \bibinfo{person}{Ziwen Jiang},
  \bibinfo{person}{Yufei Feng}, \bibinfo{person}{Binbin Hu},
  \bibinfo{person}{Kaiqi Zhao}, \bibinfo{person}{Qingwen Liu}, {and}
  \bibinfo{person}{Wenwu Ou}.} \bibinfo{year}{2020}\natexlab{}.
\newblock \showarticletitle{EdgeRec: Recommender System on Edge in Mobile
  Taobao}. In \bibinfo{booktitle}{\emph{CIKM}}. \bibinfo{pages}{2477--2484}.
\newblock


\bibitem[\protect\citeauthoryear{Guo, Liu, Kannan, Minka, Taylor, Wang, and
  Faloutsos}{Guo et~al\mbox{.}}{2009}]%
        {guo2009click}
\bibfield{author}{\bibinfo{person}{Fan Guo}, \bibinfo{person}{Chao Liu},
  \bibinfo{person}{Anitha Kannan}, \bibinfo{person}{Tom Minka},
  \bibinfo{person}{Michael Taylor}, \bibinfo{person}{Yi-Min Wang}, {and}
  \bibinfo{person}{Christos Faloutsos}.} \bibinfo{year}{2009}\natexlab{}.
\newblock \showarticletitle{Click chain model in web search}. In
  \bibinfo{booktitle}{\emph{WWW}}. \bibinfo{pages}{11--20}.
\newblock


\bibitem[\protect\citeauthoryear{Guo, Tang, Ye, Li, and He}{Guo
  et~al\mbox{.}}{2017}]%
        {guo2017deepfm}
\bibfield{author}{\bibinfo{person}{Huifeng Guo}, \bibinfo{person}{Ruiming
  Tang}, \bibinfo{person}{Yunming Ye}, \bibinfo{person}{Zhenguo Li}, {and}
  \bibinfo{person}{Xiuqiang He}.} \bibinfo{year}{2017}\natexlab{}.
\newblock \showarticletitle{DeepFM: a factorization-machine based neural
  network for CTR prediction}. In \bibinfo{booktitle}{\emph{IJCAI}}.
  \bibinfo{pages}{1725--1731}.
\newblock


\bibitem[\protect\citeauthoryear{Guo, Yu, Liu, Tang, and Zhang}{Guo
  et~al\mbox{.}}{2019}]%
        {guo2019pal}
\bibfield{author}{\bibinfo{person}{Huifeng Guo}, \bibinfo{person}{Jinkai Yu},
  \bibinfo{person}{Qing Liu}, \bibinfo{person}{Ruiming Tang}, {and}
  \bibinfo{person}{Yuzhou Zhang}.} \bibinfo{year}{2019}\natexlab{}.
\newblock \showarticletitle{PAL: a position-bias aware learning framework for
  CTR prediction in live recommender systems}. In
  \bibinfo{booktitle}{\emph{RecSys}}. \bibinfo{pages}{452--456}.
\newblock


\bibitem[\protect\citeauthoryear{He and McAuley}{He and McAuley}{2016}]%
        {he2016fusing}
\bibfield{author}{\bibinfo{person}{Ruining He} {and} \bibinfo{person}{Julian
  McAuley}.} \bibinfo{year}{2016}\natexlab{}.
\newblock \showarticletitle{Fusing similarity models with markov chains for
  sparse sequential recommendation}. In \bibinfo{booktitle}{\emph{ICDM}}.
  \bibinfo{pages}{191--200}.
\newblock


\bibitem[\protect\citeauthoryear{Hidasi, Karatzoglou, Baltrunas, and
  Tikk}{Hidasi et~al\mbox{.}}{2016}]%
        {hidasi2015session}
\bibfield{author}{\bibinfo{person}{Bal{\'a}zs Hidasi},
  \bibinfo{person}{Alexandros Karatzoglou}, \bibinfo{person}{Linas Baltrunas},
  {and} \bibinfo{person}{Domonkos Tikk}.} \bibinfo{year}{2016}\natexlab{}.
\newblock \showarticletitle{Session-based recommendations with recurrent neural
  networks}. In \bibinfo{booktitle}{\emph{ICLR}}.
\newblock


\bibitem[\protect\citeauthoryear{Hou, Hu, Zhang, and Zhao}{Hou
  et~al\mbox{.}}{2022}]%
        {hou2022core}
\bibfield{author}{\bibinfo{person}{Yupeng Hou}, \bibinfo{person}{Binbin Hu},
  \bibinfo{person}{Zhiqiang Zhang}, {and} \bibinfo{person}{Wayne~Xin Zhao}.}
  \bibinfo{year}{2022}\natexlab{}.
\newblock \showarticletitle{Core: simple and effective session-based
  recommendation within consistent representation space}. In
  \bibinfo{booktitle}{\emph{SIGIR}}. \bibinfo{pages}{1796--1801}.
\newblock


\bibitem[\protect\citeauthoryear{Hu, Wu, Zhou, Liu, Huangfu, Zhang, and
  Chen}{Hu et~al\mbox{.}}{2022}]%
        {humerit}
\bibfield{author}{\bibinfo{person}{Binbin Hu}, \bibinfo{person}{Zhengwei Wu},
  \bibinfo{person}{Jun Zhou}, \bibinfo{person}{Ziqi Liu},
  \bibinfo{person}{Zhigang Huangfu}, \bibinfo{person}{Zhiqiang Zhang}, {and}
  \bibinfo{person}{Chaochao Chen}.} \bibinfo{year}{2022}\natexlab{}.
\newblock \showarticletitle{MERIT: Learning Multi-level Representations on
  Temporal Graphs}. In \bibinfo{booktitle}{\emph{IJCAI}}.
\newblock


\bibitem[\protect\citeauthoryear{Jin, Fang, Zhang, Ren, Zhou, Xu, Yu, Wang,
  Zhu, and Gai}{Jin et~al\mbox{.}}{2020}]%
        {jin2020deep}
\bibfield{author}{\bibinfo{person}{Jiarui Jin}, \bibinfo{person}{Yuchen Fang},
  \bibinfo{person}{Weinan Zhang}, \bibinfo{person}{Kan Ren},
  \bibinfo{person}{Guorui Zhou}, \bibinfo{person}{Jian Xu},
  \bibinfo{person}{Yong Yu}, \bibinfo{person}{Jun Wang},
  \bibinfo{person}{Xiaoqiang Zhu}, {and} \bibinfo{person}{Kun Gai}.}
  \bibinfo{year}{2020}\natexlab{}.
\newblock \showarticletitle{A deep recurrent survival model for unbiased
  ranking}. In \bibinfo{booktitle}{\emph{SIGIR}}. \bibinfo{pages}{29--38}.
\newblock


\bibitem[\protect\citeauthoryear{Joachims, Swaminathan, and Schnabel}{Joachims
  et~al\mbox{.}}{2017}]%
        {joachims2017unbiased}
\bibfield{author}{\bibinfo{person}{Thorsten Joachims}, \bibinfo{person}{Adith
  Swaminathan}, {and} \bibinfo{person}{Tobias Schnabel}.}
  \bibinfo{year}{2017}\natexlab{}.
\newblock \showarticletitle{Unbiased learning-to-rank with biased feedback}. In
  \bibinfo{booktitle}{\emph{WSDM}}. \bibinfo{pages}{781--789}.
\newblock


\bibitem[\protect\citeauthoryear{Kang and McAuley}{Kang and McAuley}{2018}]%
        {kang2018self}
\bibfield{author}{\bibinfo{person}{Wang-Cheng Kang} {and}
  \bibinfo{person}{Julian McAuley}.} \bibinfo{year}{2018}\natexlab{}.
\newblock \showarticletitle{Self-attentive sequential recommendation}. In
  \bibinfo{booktitle}{\emph{ICDM}}. \bibinfo{pages}{197--206}.
\newblock


\bibitem[\protect\citeauthoryear{Lian, Zhou, Zhang, Chen, Xie, and Sun}{Lian
  et~al\mbox{.}}{2018}]%
        {lian2018xdeepfm}
\bibfield{author}{\bibinfo{person}{Jianxun Lian}, \bibinfo{person}{Xiaohuan
  Zhou}, \bibinfo{person}{Fuzheng Zhang}, \bibinfo{person}{Zhongxia Chen},
  \bibinfo{person}{Xing Xie}, {and} \bibinfo{person}{Guangzhong Sun}.}
  \bibinfo{year}{2018}\natexlab{}.
\newblock \showarticletitle{xdeepfm: Combining explicit and implicit feature
  interactions for recommender systems}. In \bibinfo{booktitle}{\emph{SIGKDD}}.
  \bibinfo{pages}{1754--1763}.
\newblock


\bibitem[\protect\citeauthoryear{Luo, Zhao, Liu, Zhuang, Wang, Xu, Fang, and
  Sheng}{Luo et~al\mbox{.}}{2020}]%
        {luocollaborative}
\bibfield{author}{\bibinfo{person}{Anjing Luo}, \bibinfo{person}{Pengpeng
  Zhao}, \bibinfo{person}{Yanchi Liu}, \bibinfo{person}{Fuzhen Zhuang},
  \bibinfo{person}{Deqing Wang}, \bibinfo{person}{Jiajie Xu},
  \bibinfo{person}{Junhua Fang}, {and} \bibinfo{person}{Victor~S Sheng}.}
  \bibinfo{year}{2020}\natexlab{}.
\newblock \showarticletitle{Collaborative Self-Attention Network for
  Session-based Recommendation}. In \bibinfo{booktitle}{\emph{IJCAI}}.
  \bibinfo{pages}{2591--2597}.
\newblock


\bibitem[\protect\citeauthoryear{Ma, Zhao, Yi, Chen, Hong, and Chi}{Ma
  et~al\mbox{.}}{2018}]%
        {ma2018modeling}
\bibfield{author}{\bibinfo{person}{Jiaqi Ma}, \bibinfo{person}{Zhe Zhao},
  \bibinfo{person}{Xinyang Yi}, \bibinfo{person}{Jilin Chen},
  \bibinfo{person}{Lichan Hong}, {and} \bibinfo{person}{Ed~H Chi}.}
  \bibinfo{year}{2018}\natexlab{}.
\newblock \showarticletitle{Modeling task relationships in multi-task learning
  with multi-gate mixture-of-experts}. In \bibinfo{booktitle}{\emph{SIGKDD}}.
  \bibinfo{pages}{1930--1939}.
\newblock


\bibitem[\protect\citeauthoryear{Minh, Niyogi, and Yao}{Minh
  et~al\mbox{.}}{2006}]%
        {minh2006mercer}
\bibfield{author}{\bibinfo{person}{Ha~Quang Minh}, \bibinfo{person}{Partha
  Niyogi}, {and} \bibinfo{person}{Yuan Yao}.} \bibinfo{year}{2006}\natexlab{}.
\newblock \showarticletitle{Mercer’s theorem, feature maps, and smoothing}.
  In \bibinfo{booktitle}{\emph{COLT}}. \bibinfo{pages}{154--168}.
\newblock


\bibitem[\protect\citeauthoryear{Omi, Ueda, and Aihara}{Omi
  et~al\mbox{.}}{2019}]%
        {omi2019fully}
\bibfield{author}{\bibinfo{person}{Takahiro Omi}, \bibinfo{person}{Naonori
  Ueda}, {and} \bibinfo{person}{Kazuyuki Aihara}.}
  \bibinfo{year}{2019}\natexlab{}.
\newblock \showarticletitle{Fully neural network based model for general
  temporal point processes}. In \bibinfo{booktitle}{\emph{NIPS}}.
  \bibinfo{pages}{2120--2129}.
\newblock


\bibitem[\protect\citeauthoryear{Qin, Chen, Metzler, Noh, Qin, and Wang}{Qin
  et~al\mbox{.}}{2020}]%
        {qin2020attribute}
\bibfield{author}{\bibinfo{person}{Zhen Qin}, \bibinfo{person}{Suming~J Chen},
  \bibinfo{person}{Donald Metzler}, \bibinfo{person}{Yongwoo Noh},
  \bibinfo{person}{Jingzheng Qin}, {and} \bibinfo{person}{Xuanhui Wang}.}
  \bibinfo{year}{2020}\natexlab{}.
\newblock \showarticletitle{Attribute-based propensity for unbiased learning in
  recommender systems: Algorithm and case studies}. In
  \bibinfo{booktitle}{\emph{SIGKDD}}. \bibinfo{pages}{2359--2367}.
\newblock


\bibitem[\protect\citeauthoryear{Rendle, Freudenthaler, and
  Schmidt-Thieme}{Rendle et~al\mbox{.}}{2010}]%
        {rendle2010factorizing}
\bibfield{author}{\bibinfo{person}{Steffen Rendle}, \bibinfo{person}{Christoph
  Freudenthaler}, {and} \bibinfo{person}{Lars Schmidt-Thieme}.}
  \bibinfo{year}{2010}\natexlab{}.
\newblock \showarticletitle{Factorizing personalized markov chains for
  next-basket recommendation}. In \bibinfo{booktitle}{\emph{WWW}}.
  \bibinfo{pages}{811--820}.
\newblock


\bibitem[\protect\citeauthoryear{Tan, Xu, and Liu}{Tan et~al\mbox{.}}{2016}]%
        {tan2016improved}
\bibfield{author}{\bibinfo{person}{Yong~Kiam Tan}, \bibinfo{person}{Xinxing
  Xu}, {and} \bibinfo{person}{Yong Liu}.} \bibinfo{year}{2016}\natexlab{}.
\newblock \showarticletitle{Improved recurrent neural networks for
  session-based recommendations}. In \bibinfo{booktitle}{\emph{RecSys
  Workshop}}. \bibinfo{pages}{17--22}.
\newblock


\bibitem[\protect\citeauthoryear{Tang and Wang}{Tang and Wang}{2018}]%
        {tang2018personalized}
\bibfield{author}{\bibinfo{person}{Jiaxi Tang} {and} \bibinfo{person}{Ke
  Wang}.} \bibinfo{year}{2018}\natexlab{}.
\newblock \showarticletitle{Personalized top-n sequential recommendation via
  convolutional sequence embedding}. In \bibinfo{booktitle}{\emph{WSDM}}.
  \bibinfo{pages}{565--573}.
\newblock


\bibitem[\protect\citeauthoryear{Vaswani, Shazeer, Parmar, Uszkoreit, Jones,
  Gomez, Kaiser, and Polosukhin}{Vaswani et~al\mbox{.}}{2017}]%
        {vaswani2017attention}
\bibfield{author}{\bibinfo{person}{Ashish Vaswani}, \bibinfo{person}{Noam
  Shazeer}, \bibinfo{person}{Niki Parmar}, \bibinfo{person}{Jakob Uszkoreit},
  \bibinfo{person}{Llion Jones}, \bibinfo{person}{Aidan~N Gomez},
  \bibinfo{person}{Lukasz Kaiser}, {and} \bibinfo{person}{Illia Polosukhin}.}
  \bibinfo{year}{2017}\natexlab{}.
\newblock \showarticletitle{Attention is all you need}. In
  \bibinfo{booktitle}{\emph{NIPS}}. \bibinfo{pages}{5998--6008}.
\newblock


\bibitem[\protect\citeauthoryear{Wang, Golbandi, Bendersky, Metzler, and
  Najork}{Wang et~al\mbox{.}}{2018}]%
        {wang2018position}
\bibfield{author}{\bibinfo{person}{Xuanhui Wang}, \bibinfo{person}{Nadav
  Golbandi}, \bibinfo{person}{Michael Bendersky}, \bibinfo{person}{Donald
  Metzler}, {and} \bibinfo{person}{Marc Najork}.}
  \bibinfo{year}{2018}\natexlab{}.
\newblock \showarticletitle{Position bias estimation for unbiased learning to
  rank in personal search}. In \bibinfo{booktitle}{\emph{WSDM}}.
  \bibinfo{pages}{610--618}.
\newblock


\bibitem[\protect\citeauthoryear{Wu, He, Wang, Zhang, and Wang}{Wu
  et~al\mbox{.}}{2021}]%
        {wu2021survey}
\bibfield{author}{\bibinfo{person}{Le Wu}, \bibinfo{person}{Xiangnan He},
  \bibinfo{person}{Xiang Wang}, \bibinfo{person}{Kun Zhang}, {and}
  \bibinfo{person}{Meng Wang}.} \bibinfo{year}{2021}\natexlab{}.
\newblock \showarticletitle{A Survey on Neural Recommendation: From
  Collaborative Filtering to Content and Context Enriched Recommendation}.
\newblock \bibinfo{journal}{\emph{arXiv preprint arXiv:2104.13030}}
  (\bibinfo{year}{2021}).
\newblock


\bibitem[\protect\citeauthoryear{Xiao, Ye, He, Zhang, Wu, and Chua}{Xiao
  et~al\mbox{.}}{2017}]%
        {xiao2017attentional}
\bibfield{author}{\bibinfo{person}{Jun Xiao}, \bibinfo{person}{Hao Ye},
  \bibinfo{person}{Xiangnan He}, \bibinfo{person}{Hanwang Zhang},
  \bibinfo{person}{Fei Wu}, {and} \bibinfo{person}{Tat-Seng Chua}.}
  \bibinfo{year}{2017}\natexlab{}.
\newblock \showarticletitle{Attentional factorization machines: Learning the
  weight of feature interactions via attention networks}. In
  \bibinfo{booktitle}{\emph{IJCAI}}. \bibinfo{pages}{3119--3125}.
\newblock


\bibitem[\protect\citeauthoryear{Xu, Zhao, Liu, Xu, S.~Sheng, Cui, Zhou, and
  Xiong}{Xu et~al\mbox{.}}{2019b}]%
        {xu2019recurrent}
\bibfield{author}{\bibinfo{person}{Chengfeng Xu}, \bibinfo{person}{Pengpeng
  Zhao}, \bibinfo{person}{Yanchi Liu}, \bibinfo{person}{Jiajie Xu},
  \bibinfo{person}{Victor S~Sheng S.~Sheng}, \bibinfo{person}{Zhiming Cui},
  \bibinfo{person}{Xiaofang Zhou}, {and} \bibinfo{person}{Hui Xiong}.}
  \bibinfo{year}{2019}\natexlab{b}.
\newblock \showarticletitle{Recurrent convolutional neural network for
  sequential recommendation}. In \bibinfo{booktitle}{\emph{WWW}}.
  \bibinfo{pages}{3398--3404}.
\newblock


\bibitem[\protect\citeauthoryear{Xu, Ruan, Kumar, Korpeoglu, and Achan}{Xu
  et~al\mbox{.}}{2019a}]%
        {xu2019self}
\bibfield{author}{\bibinfo{person}{Da Xu}, \bibinfo{person}{Chuanwei Ruan},
  \bibinfo{person}{Sushant Kumar}, \bibinfo{person}{Evren Korpeoglu}, {and}
  \bibinfo{person}{Kannan Achan}.} \bibinfo{year}{2019}\natexlab{a}.
\newblock \showarticletitle{Self-attention with functional time representation
  learning}. In \bibinfo{booktitle}{\emph{NIPS}}.
  \bibinfo{pages}{15915–--15925}.
\newblock


\bibitem[\protect\citeauthoryear{Yan, Cheng, Kang, Wan, and McAuley}{Yan
  et~al\mbox{.}}{2019}]%
        {yan2019cosrec}
\bibfield{author}{\bibinfo{person}{An Yan}, \bibinfo{person}{Shuo Cheng},
  \bibinfo{person}{Wang-Cheng Kang}, \bibinfo{person}{Mengting Wan}, {and}
  \bibinfo{person}{Julian McAuley}.} \bibinfo{year}{2019}\natexlab{}.
\newblock \showarticletitle{CosRec: 2D convolutional neural networks for
  sequential recommendation}. In \bibinfo{booktitle}{\emph{CIKM}}.
  \bibinfo{pages}{2173--2176}.
\newblock


\bibitem[\protect\citeauthoryear{Ye, Wang, Chen, Wang, Qin, and Yin}{Ye
  et~al\mbox{.}}{2020}]%
        {ye2020time}
\bibfield{author}{\bibinfo{person}{Wenwen Ye}, \bibinfo{person}{Shuaiqiang
  Wang}, \bibinfo{person}{Xu Chen}, \bibinfo{person}{Xuepeng Wang},
  \bibinfo{person}{Zheng Qin}, {and} \bibinfo{person}{Dawei Yin}.}
  \bibinfo{year}{2020}\natexlab{}.
\newblock \showarticletitle{Time Matters: Sequential Recommendation with
  Complex Temporal Information}. In \bibinfo{booktitle}{\emph{SIGIR}}.
  \bibinfo{pages}{1459--1468}.
\newblock


\bibitem[\protect\citeauthoryear{Zhao, Hong, Wei, Chen, Nath, Andrews,
  Kumthekar, Sathiamoorthy, Yi, and Chi}{Zhao et~al\mbox{.}}{2019}]%
        {zhao2019recommending}
\bibfield{author}{\bibinfo{person}{Zhe Zhao}, \bibinfo{person}{Lichan Hong},
  \bibinfo{person}{Li Wei}, \bibinfo{person}{Jilin Chen},
  \bibinfo{person}{Aniruddh Nath}, \bibinfo{person}{Shawn Andrews},
  \bibinfo{person}{Aditee Kumthekar}, \bibinfo{person}{Maheswaran
  Sathiamoorthy}, \bibinfo{person}{Xinyang Yi}, {and} \bibinfo{person}{Ed
  Chi}.} \bibinfo{year}{2019}\natexlab{}.
\newblock \showarticletitle{Recommending what video to watch next: a multitask
  ranking system}. In \bibinfo{booktitle}{\emph{RecSys}}.
  \bibinfo{pages}{43--51}.
\newblock


\bibitem[\protect\citeauthoryear{Zhou, Mou, Fan, Pi, Bian, Zhou, Zhu, and
  Gai}{Zhou et~al\mbox{.}}{2019}]%
        {zhou2019deep}
\bibfield{author}{\bibinfo{person}{Guorui Zhou}, \bibinfo{person}{Na Mou},
  \bibinfo{person}{Ying Fan}, \bibinfo{person}{Qi Pi}, \bibinfo{person}{Weijie
  Bian}, \bibinfo{person}{Chang Zhou}, \bibinfo{person}{Xiaoqiang Zhu}, {and}
  \bibinfo{person}{Kun Gai}.} \bibinfo{year}{2019}\natexlab{}.
\newblock \showarticletitle{Deep interest evolution network for click-through
  rate prediction}. In \bibinfo{booktitle}{\emph{AAAI}}.
  \bibinfo{pages}{5941--5948}.
\newblock


\bibitem[\protect\citeauthoryear{Zhou, Zhu, Song, Fan, Zhu, Ma, Yan, Jin, Li,
  and Gai}{Zhou et~al\mbox{.}}{2018}]%
        {zhou2018deep}
\bibfield{author}{\bibinfo{person}{Guorui Zhou}, \bibinfo{person}{Xiaoqiang
  Zhu}, \bibinfo{person}{Chenru Song}, \bibinfo{person}{Ying Fan},
  \bibinfo{person}{Han Zhu}, \bibinfo{person}{Xiao Ma},
  \bibinfo{person}{Yanghui Yan}, \bibinfo{person}{Junqi Jin},
  \bibinfo{person}{Han Li}, {and} \bibinfo{person}{Kun Gai}.}
  \bibinfo{year}{2018}\natexlab{}.
\newblock \showarticletitle{Deep interest network for click-through rate
  prediction}. In \bibinfo{booktitle}{\emph{SIGKDD}}.
  \bibinfo{pages}{1059--1068}.
\newblock


\bibitem[\protect\citeauthoryear{Zhou, Li, Zhao, Chen, Li, Yang, Cui, Yu, Chen,
  Ding, et~al\mbox{.}}{Zhou et~al\mbox{.}}{2017}]%
        {zhou2017kunpeng}
\bibfield{author}{\bibinfo{person}{Jun Zhou}, \bibinfo{person}{Xiaolong Li},
  \bibinfo{person}{Peilin Zhao}, \bibinfo{person}{Chaochao Chen},
  \bibinfo{person}{Longfei Li}, \bibinfo{person}{Xinxing Yang},
  \bibinfo{person}{Qing Cui}, \bibinfo{person}{Jin Yu}, \bibinfo{person}{Xu
  Chen}, \bibinfo{person}{Yi Ding}, {et~al\mbox{.}}}
  \bibinfo{year}{2017}\natexlab{}.
\newblock \showarticletitle{Kunpeng: Parameter server based distributed
  learning systems and its applications in alibaba and ant financial}. In
  \bibinfo{booktitle}{\emph{SIGKDD}}. \bibinfo{pages}{1693--1702}.
\newblock


\bibitem[\protect\citeauthoryear{Zhu, Li, Liao, Wang, Guan, Liu, and Cai}{Zhu
  et~al\mbox{.}}{2017}]%
        {zhu2017next}
\bibfield{author}{\bibinfo{person}{Yu Zhu}, \bibinfo{person}{Hao Li},
  \bibinfo{person}{Yikang Liao}, \bibinfo{person}{Beidou Wang},
  \bibinfo{person}{Ziyu Guan}, \bibinfo{person}{Haifeng Liu}, {and}
  \bibinfo{person}{Deng Cai}.} \bibinfo{year}{2017}\natexlab{}.
\newblock \showarticletitle{What to Do Next: Modeling User Behaviors by
  Time-LSTM.}. In \bibinfo{booktitle}{\emph{IJCAI}}.
  \bibinfo{pages}{3602--3608}.
\newblock


\bibitem[\protect\citeauthoryear{Zhu, Chen, Minka, Zhu, and Chen}{Zhu
  et~al\mbox{.}}{2010}]%
        {zhu2010novel}
\bibfield{author}{\bibinfo{person}{Zeyuan~Allen Zhu}, \bibinfo{person}{Weizhu
  Chen}, \bibinfo{person}{Tom Minka}, \bibinfo{person}{Chenguang Zhu}, {and}
  \bibinfo{person}{Zheng Chen}.} \bibinfo{year}{2010}\natexlab{}.
\newblock \showarticletitle{A novel click model and its applications to online
  advertising}. In \bibinfo{booktitle}{\emph{WSDM}}. \bibinfo{pages}{321--330}.
\newblock


\bibitem[\protect\citeauthoryear{Zimdars, Chickering, and Meek}{Zimdars
  et~al\mbox{.}}{2001}]%
        {zimdars2013using}
\bibfield{author}{\bibinfo{person}{Andrew Zimdars},
  \bibinfo{person}{David~Maxwell Chickering}, {and}
  \bibinfo{person}{Christopher Meek}.} \bibinfo{year}{2001}\natexlab{}.
\newblock \showarticletitle{Using temporal data for making recommendations}. In
  \bibinfo{booktitle}{\emph{UAI}}. \bibinfo{pages}{580--588}.
\newblock


\end{thebibliography}
